\documentclass[twocolumn]{aastex631}
\usepackage{hyperref}

\usepackage{comment}
\usepackage{amsmath}
\usepackage{multirow}
\usepackage{orcidlink}
\usepackage{cancel}

\newcommand{\afterglowpy}{{\tt afterglowpy}}

\newcommand{\sm}[1]{\textcolor{black}{#1}} 
\newcommand{\gsr}[1]{\textcolor{black}{#1}} 
\newcommand{\mm}[1]{\textcolor{black}{#1}} 

\newcommand{\mmfix}[1]{\textcolor{black}{#1}} 
\newcommand{\gsrfix}[1]{\textcolor{black}{#1}} 
\newcommand{\smn}[1]{\textcolor{black}{#1}} 

\newcommand{\lb}{\left(}
\newcommand{\rb}{\right)}
\newcommand{\lp}{\left[}
\newcommand{\rp}{\right]}
\newcommand{\lc}{\left\lbrace}
\newcommand{\rc}{\right\rbrace}

\newcommand{\degree}{^{\circ}}                  

\newcommand{\post}{p}
\newcommand{\like}{\mathcal{L}}
\newcommand{\prio}{\pi}
\newcommand{\sem}{_{\text{EM}}}
\newcommand{\sgw}{_{\text{GW}}}

\newcommand{\tjn}{\theta_{\text{JN}}}
\newcommand{\dlum}{d_L}
\newcommand{\avi}{\theta_{\text{vi}}}

\newcommand{\tmis}{\theta_{\text{m}}}            

\newcommand{\lamH}{ \lambda_{H_0} }

\received{June 17, 2024}
\revised{October 7, 2024}
\accepted{October 24, 2024}

\submitjournal{ApJL}

\shorttitle{Systematics in Bright sirens cosmology}

\shortauthors{M\"{u}ller, Mukherjee, Ryan}

\begin{document}

\title{Be careful in multi-messenger inference of the Hubble constant: A path forward for robust inference}

\author[0000-0003-4989-6154]{Michael M\"{u}ller}
\affiliation{Institute of Physics, University of Greifswald, D-17489 Greifswald, Germany}
\affiliation{Perimeter Institute for Theoretical Physics, Waterloo, Ontario N2L 2Y5, Canada}
\affiliation{Department of Physics, University of Guelph, Guelph, Ontario N1G 2W1, Canada}

\author[0000-0002-3373-5236]{Suvodip Mukherjee}
\affiliation{Department of Astronomy and Astrophysics, Tata Institute of Fundamental Research, 1, Homi Bhabha Road, Mumbai, 400005, India}

\author[0000-0001-9068-7157]{Geoffrey Ryan}
\affiliation{Perimeter Institute for Theoretical Physics, Waterloo, Ontario N2L 2Y5, Canada}

\correspondingauthor{Michael M\"{u}ller \\
michael.mueller@uni-greifswald.de}




\begin{abstract}
\gsr{Multi-messenger observations of coalescing binary neutron stars (BNSs) \sm{using gravitational wave (GW) and electromagnetic (EM) wave signals} are a direct probe of the expansion history of the universe and carry the potential to shed light on the disparity between low- and high-redshift measurements of the Hubble constant $H_0$.}
To measure the value of $H_0$ with such observations requires pristine inference of the luminosity distance and the true source redshift with minimal impact from systematics.
In this analysis, we carry out joint inference on mock GW signals and their EM afterglows from BNS coalescences and find that the inclination angle inferred from the afterglow light curve and apparent superluminal motion can be precise, but need not be accurate and is subject to systematic uncertainty \mm{that could be as large as $1.5\sigma$}.
This produces a disparity between the EM and GW inferred inclination angles, which if not carefully treated when combining observations can bias the inferred value of $H_0$.
\mm{We also find that already small misalignments of $3^{\circ}-6^{\circ}$ between the inherent system inclinations for the GW and EM emission can bias the inference by $\mathcal{O}(1-2\sigma)$ if not taken into account.}
As multi-messenger BNS observations are rare, we must make the most out of a small number of events and harness
the increased precision, while avoiding a reduced accuracy.
We demonstrate how to mitigate these potential sources of bias by jointly inferring the mismatch between the GW- and EM-based inclination angles and $H_0$. 
\end{abstract}

\keywords{gravitational waves --- cosmology: distance scale }

\section{Introduction} \label{sec:intro}

The \textit{bright sirens} method for measuring the Hubble flow in the \gsr{local universe accessible to gravitational wave (GW) detectors was proposed by \citet{schutz_determining_1986} based on the robust prediction by  General Relativity (GR) of the luminosity and time evolution of a binary neutron star (BNS) coalescence. }
\gsr{It has been validated at the single-event level by the multi-messenger observation of GW170817, which provided luminosity distance and redshift information independently \citep[e.g.][]{abbott_multi-messenger_2017} and }produced a \mmfix{$\sim 15\%$ measurement of the Hubble parameter $H_0$, see \cite{abbott_gravitational-wave_2017}}. \smn{This measurement of the Hubble parameter is further improved to a $\sim 7\%$ measurement by including the observation of radio jet and improved peculiar velocity (see \citealt{abbott_gravitational-wave_2017,hotokezaka_hubble_2018, mukherjee_velocity_2021}).} 
Though such measurements are robust theoretically, in practice multiple astrophysical uncertainties such as inclination angle \citep{nissanke_exploring_2010}, \mmfix{peculiar velocity \citep{howlett_standard_2020,mukherjee_velocity_2021, nicolaou_impact_2020,Nimonkar:2023pyt}}, waveform uncertainty \citep{Kunert:2024xqb}, non-stationary noise \citep{PhysRevD.106.043504}, etc. can obscure the inference of $H_0$ if not taken into account correctly. \sm{To accurately infer the value of the Hubble constant from GW data and shed light on the ongoing tension on the true value of the Hubble constant (see \citealt{Abdalla:2022yfr} for a review), it is important to mitigate the influence of any possible systematic due to astrophysical and instrument uncertainties.}

In order to mitigate these additional uncertainties, combining GW and electromagnetic (EM) observations can be beneficial. For a \gsr{BNS} or black hole-neutron star (BHNS) merger the remnant, if matter is still present, is expected to source a highly relativistic, collimated matter outflow (jet) from its poles, which produces a short gamma-ray burst (sGRB, $\mathcal{O}(1s)$) 
and a long-lived transient spanning all wavelengths ($\mathcal{O}(\text{years})$) \gsr{called the \emph{afterglow}} (see \citealt{berger_short-duration_2014,nakar_afterglow_2021,abbott_multi-messenger_2017}).
Besides allowing for the identification of a host galaxy, and through that a
redshift, the EM counterpart can also mitigate a degeneracy between the
luminosity distance \mm{($\dlum$)} and the system orientation in the GW analysis, which limits the precision
of the $\dlum$ measurement \mmfix{\citep[e.g.][]{nissanke_exploring_2010,schutz_networks_2011}.}

\begin{figure}
\centering
\includegraphics[trim=0.1cm 1.0cm 0 0, clip,scale=0.6]{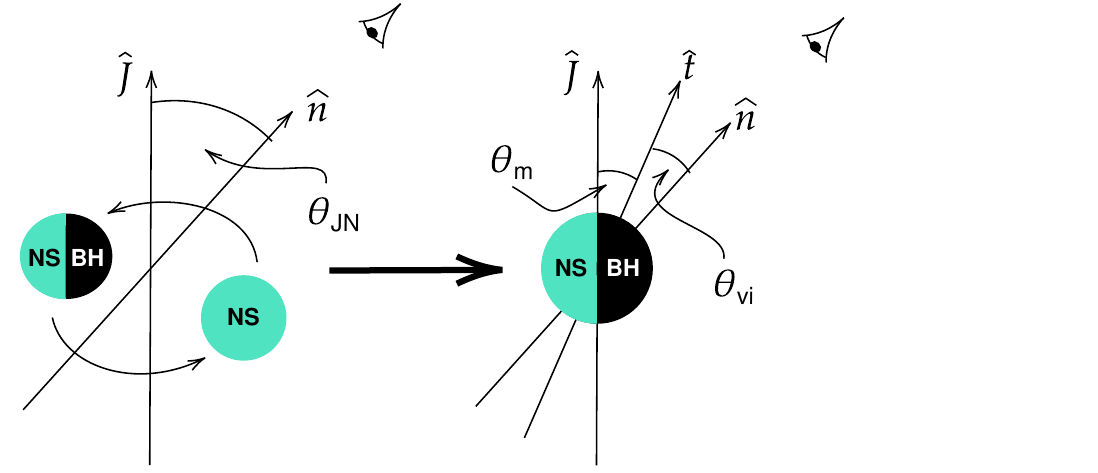}
\caption{
    Sketch for the definition of the system inclination $\tjn$ during the inspiral/merger on the
    left and for the definition of the jet viewing angle $\avi$ and misalignment angle $\tmis$
    for the post-merger remnant. The line of sight for the observer is indicated with the vector 
    $\hat{n}$.
} \label{fig:sketch}
\end{figure}

The angle between the observer line of sight $\hat{n}$ and the orbital angular momentum axis $\hat{J}$
(see Fig. \ref{fig:sketch}) is generally referred to as the system inclination, $\tjn$. Since the jet comes from a polar outflow of the remnant, $\tjn$ is expected to be correlated 
with the jet inclination\footnote{
    Note that the polar axis of the remnant, i.e. the total angular
    \mm{momentum axis post-merger might not correspond to the orbital angular momentum axis pre-merger, 
    e.g. \cite{dietrich_numerical_2018,chaurasia_gravitational_2020}.}
}.
While (non-)observation of the prompt $\gamma$-ray emission can provide a
weak \mm{(lower) upper bound} on the viewing angle due to its highly collimated nature,
the light curve of the afterglow, which is also detectable further off-axis, is more sensitive to the viewing angle $\avi$ (which is defined as the angle between the jets direction of propagation $\hat{t}$ and the observers' line of sight $\hat{n}$, see
Fig. \ref{fig:sketch}). In particular, the initial rising slope of the light curve
(for bright signals) and the jet break \mm{contain orientation information} and  allow
extraction of the viewing angle (e.g. \citealt{nakar_afterglow_2021,ryan_modelling_2023}). 
The constraint
from the lightcurve is actually on a combination of the viewing angle and opening
angle of the jet and this degeneracy curtails the precision of the measurement
(\citealt{ryan_gamma-ray_2020,takahashi_inverse_2020,nakar_afterglow_2021}). A possible way to break the degeneracy in the light curve data is through astrometry performed on the radio image of the afterglow, 
which can reveal a lateral \gsr{apparent superluminal} motion of the afterglow centroid
that is \gsr{highly} sensitive to the viewing angle \footnote{
    \cite{nakar_afterglow_2021,gianfagna_joint_2023} point out that data on the centroid
    motion alone constrains a combination of jet-viewing and -opening angle as well and
    therefore is also not able to infer the former precisely, unless combined with the
    light curve data.
},
and thus lead to a $10\%$-level measurement
of the viewing angle, see e.g. \cite{mooley_superluminal_2018}. \mm{Such a constraint for the
viewing angle, combined with assumptions about its correlation with the system inclination}, can limit or
break the degeneracy between $\tjn$ and $\dlum$ and improve the precision of the
$\dlum$-$H_0$-measurement by a factor of two 
\citep[][]{nissanke_exploring_2010,chen_viewing_2019}. However, by including EM data, the clear dependence of the inferred $\dlum$
on GR alone is lost
and one has to make sure that the additional precision at the single-event
level \mm{is not accompanied by a reduced accuracy in the inference.}

\mm{In particular, the} interpretation of the photometry and astrometry data for the jet afterglow
and its connection to the properties of the merging 
compact binary system \mm{requires the assumption of a model that
describes the emission from the jet afterglow}. However, the
emission model for the afterglow that is fit to the EM data to constrain
$\avi$ is strongly dependent on the assumed jet structure and geometry.
Understanding the latter based on the properties of the merging binary, \gsr{unfortunately},
is presently not feasible due to the computational challenges involved in
the physical processes at vastly different scales that govern the merger and
subsequent remnant evolution with jet formation (\citealt{combi_jets_2023,kiuchi_large-scale_2024,most_flares_2023} for BNS
and \citealt{most_accretion_2021,hayashi_general-relativistic_2022}
for BHNS
)
, and then jet propagation \citep{fernandez_long-term_2019,kathirgamaraju_em_2019}
and non-thermal emission \citep[e.g.][]{granot_shape_2002}.
Thus, one has to assume a particular model for the jet structure to fit the
light curve and centroid motion, which cannot be constrained by knowledge about
the inspiralling binary from the GW data. We refer to the systematic uncertainty 
in the inference of $H_0$, if the inference of $\avi$ is incorrect due to a 
theoretical modeling choice, as \textit{model} bias.

Furthermore, while the GW and EM emission
share the luminosity distance as a common parameter, it is necessary to specify
a relation between the viewing angle for the jet and the system inclination for
the inspiralling binary.
It is often assumed that
the jet axis is aligned with the \mm{total angular momentum of the post-merger system
and that this in turn does not deviate much from the
orbital angular momentum of the merging system, see e.g.
\cite{abbott_gravitational_2017,chen_viewing_2019,farah_counting_2020}, which
allows for a direct identification of the two parameters, when the
posteriors on $\tjn$ and $\avi$ are combined}. However, there is presently no
theoretical framework for confirming this assumption
and in general a misalignment of $\mathcal{O}(\text{few}\degree)$ between the \mm{three}
axes might be present due to remnant properties 
\citep[e.g.][]{stone_pulsations_2013,li__quasiperiodic_2023,hayashi_general-relativistic_2022}
or collimation in an anisotropic medium \citep[e.g.][]{nagakura_jet_2014}. 
\gsr{We define the line-of-sight misalignment angle as $\tmis = \tjn - \avi$}\footnote{
    The jet is a polar outflow from both poles and consequently scenario with
    $\avi > \pi/2$ has an equivalent and indistinguishable scenario with
    $\avi < \pi/2$ and it is therefore sufficient to consider the range
    $\avi \in [0,\pi/2]$. Consequently, when we want to compare this with the
    system inclination, we also need to project the former into this range,
    using the map $\avi\lb\tjn\rb = \frac{\pi}{2}-\vert\tjn-\frac{\pi}{2}\vert$,
    see \cite{abbott_gravitational_2017}.  \gsr{The line-of-sight misalignment $\tmis$ is a lower bound to the full misalignment $\cos^{-1}( \hat{J} \cdot \hat{t} )$.  The rest of the misalignment will show itself as a difference in orientations of the jet and GW systems on the sky, which we do not pursue further in this work.}
}
, see Fig. \ref{fig:sketch}, and refer to systematic deviations in $H_0$ due to 
an intrinsic mismatch between the jet angle and inclination angle
($\tmis^{(\text{true})}\neq0$) as \textit{misalignment} bias. 

\begin{figure}
\centering
\includegraphics[trim=1.0cm 0.5cm 0 2.0cm, clip,scale=0.38]{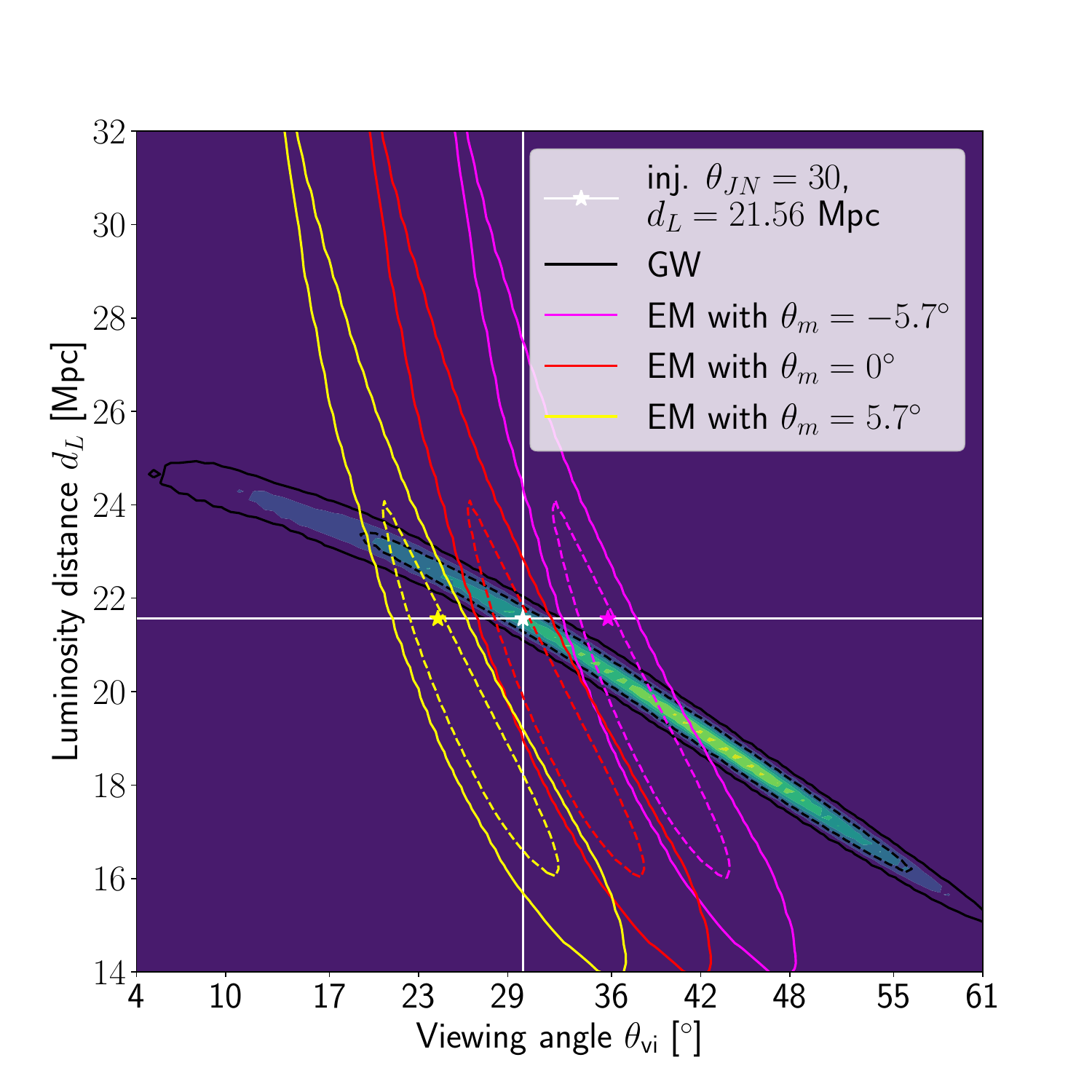}
\caption{The GW (heat-map and black contours) and EM posteriors (red contours) in the \mm{$\dlum$-$\avi$-plane},
for injected/true parameters $\tjn=30\degree$ and $\dlum=21.56$ Mpc (marked in white).
The dashed and solid contours indicate the $1\sigma$ and $2\sigma$ \mmfix{highest posterior density (HPD)} credible regions.
The yellow and magenta contours show two EM posteriors
with a positive or negative angular shift of $5.7\degree$ between the jet angle and system inclination
(the injected/true $\avi$ is indicated with stars).
} \label{fig:tmisEffect}
\end{figure}

The effect of \textit{model} and \textit{misalignment} biases is similar in
the sense that both will shift the EM posterior with respect to the GW posterior
and consequently, the region of joint posterior support, as shown in Fig.
\ref{fig:tmisEffect}, and a failure to take this into account in the joint
inference will bias the combined posterior for $\dlum$. In particular,
Fig. \ref{fig:tmisEffect} shows that a positive/negative misalignment angle, which
is equivalent to an under-/overestimation of the viewing angle in the \textit{model}
bias scenario will bias the posterior support towards larger/smaller $\dlum$
due to the anti-correlation of the GW posterior in the $\dlum$-$\tjn$-plane. 

Previously, the presence of a \textit{model bias} has been studied in great detail
for the EM counterpart of GW170817,
\cite{mooley_superluminal_2018,doctor_thunder_2020,nakar_afterglow_2021,gianfagna_joint_2023,gianfagna_potential_2024,govreen-segal_analytic_2023}.
While \mm{these authors}
study the systematic based on GW170817 and/or on future similar events\footnote{
    This is justified by the theoretical expectation that the majority of mergers
    observed in the future will have moderate inclinations (\citealt{schutz_networks_2011}),
    similar to GW170817.
} 
with detailed EM modeling, the more general analysis of \cite{chen_systematic_2020}
uses Gaussian estimates for EM likelihoods\footnote{
    Furthermore, they use a fixed bias for all events, while we are trying to quantify
    the actual bias through modeling choices, which can vary from event to event and also
    bias in different directions, we need to point out this difference more clearly.
}. 
Consequently, a systematic exploration of the \textit{model bias} for systems that differ from GW170817 that employs realistic EM inference models \mm{and mock data} has been missing so far.


\mm{This letter points out}, that the misalignment between the jet direction inferred from EM observations and GW observations can be a potential hindrance in achieving accurate $H_0$ \mm{at the single-event level} and can be the ultimate bottleneck when the statistical uncertainty is negligible. By performing joint cosmological inference on a suite of \mm{realistic} mock data simulations, we show that due to a positive/negative value of the misalignment angle, 
there can be an \mm{under-/overestimation of} $H_0$. 
We propose a Bayesian technique to mitigate this effect in a multi-messenger analysis where light-curve, centroid motion of the jet, and GW data are used to infer the value of $H_0$. 

\section{Methods}

To study the impact of potential \textit{model} and \textit{misalignment} biases on the
joint GW+EM inference, we generate a series of mock observations in the GW and EM \mm{sectors} that
are consistent with an equal mass BNS,
assuming different system inclinations and jet viewing angles in the range $0\degree-80\degree$, 
which allows us to
combine posteriors from both sources that either have an intrinsic misalignment between
the jet and orbital angular momentum axes ($\tmis^{(\text{true})} \neq 0$) or not.

\gsrfix{We restrict the present analysis to BNS systems for simplicity.  While BHNS systems may differ in the details of the jet launching mechanism and structure, if they launch a relativistic jet we expect broadly similar phenomenology to a BNS: a GW inspiral with $\dlum$-$\avi$-degeneracy and a non-thermal EM synchrotron afterglow (see also  \cite{ruiz_effects_2019}).}
\mmfix{While the asymmetries introduced by unequal masses or misaligned component spins
have, in principle, the capacity to lead to a misaligned jet, a more complete theoretical
understanding of the jet formation is required to leverage such correlations.  Consequently, as a first step we only consider equal mass binaries in this analysis.}
We consider a Flat Lambda-Cold Dark Matter (LCDM) cosmology \mmfix{\citep[see][]{mo2010galaxy}} with $H_0^{(\text{true})} = 70$ km s$^{-1}$ Mpc$^{-1}$ \mmfix{, consistent with \cite{abbott_gravitational-wave_2017},}
and we fix the redshift to 
$z^{(\text{true})}=0.005$, which corresponds to a luminosity distance of 
$\dlum^{(\text{true})}  = 21.56$ Mpc,
since we will not focus on the peculiar velocity bias.
This is an unrealistically small distance (for comparison GW170817 was observed at around $40$ Mpc, see
\citealt{abbott_gravitational-wave_2017}) and it will lead to super-ideal posteriors,
however, we investigate the biases under such optimal conditions to clearly
distinguish between statistical fluctuations (which should be strongly suppressed in
the considered loud signals) from systematic deviations. We shall use a particular jet
structure to generate all of the events, see Sec. \ref{sec:em_mock} for details. Below
we outline the statistical framework that is used for the joint GW+EM inference of the
Hubble parameter, as well as the methods with which we generated our mock observations
in both sectors.

\subsection{Bayesian Framework} \label{sec:bayes}

From Bayes' theorem \citep{1763RSPT...53..370B}, we obtain the joint Hubble posterior through
\begin{equation}
	\post\lb H_0 \vert D,\lamH\rb 
	\propto \prio\lb H_0\vert\lamH \rb \like\lb D \vert H_0,\lamH \rb, \label{eqn:postH0}
\end{equation}
where $D = \lc D\sem,D\sgw\rc$ is the collection of all relevant 
datasets from the EM and GW sectors respectively, \mm{and $\lamH$ represents the prior/model
assumptions under which the inference is performed}. Assuming that the measurements in
the sub-sectors are independent and that the data only depends on the event parameters
$\theta$, the data likelihood for the joint inference model is
\begin{equation}
	\like\lb D \vert H_0, \lamH\rb
	\! = \! \int \! \! \! d\theta \like\lb D\sem\vert \theta\rb \like\lb D\sgw\vert \theta\rb 
	\prio\lb \theta\vert H_0, \lamH\rb,
\end{equation}
where $\prio\lb \theta\vert H_0, \lamH\rb$ represents the prior assumptions for the
analysis of the combined data.
In this work we are 
interested in combining posterior knowledge about the redshift $z$, the luminosity
distance $\dlum$, the inclination of the GW source $\tjn$ and the viewing angle for
the post-merger jet $\avi$ to learn about $H_0$, and both observation channels have
access to different, but overlapping subsets of this parameter space. As discussed
in Sec. \ref{sec:gw_mock} and \ref{sec:em_mock} the respective models for the jet \mm{afterglow}
and the GW emission have a much higher parameter space with complex correlations, however,
we shall marginalize the full posteriors to the relevant subspace 
$\theta = \theta\sem \cup \theta\sgw$, with
$\theta\sem = \lc z, \avi, \dlum\rc$ and $\theta\sgw = \lc \dlum,\tjn\rc$.
In the choice of the $\lamH$-prior for the joint inference, we assume the 
general scenario where the jet viewing angle and system inclination differ by a misalignment
angle $\tmis$, whose value is determined by an apriori assumption in $\lamH$.
Furthermore, there is a mismatch in the parameter range of $\avi$ and $\tjn$,
since the reflection symmetry of the jet about the spin plane causes the scenarios
$\avi$ and $\avi + \frac{\pi}{2}$ to be indistinguishable. Thus, the GW inclination parameter
will always be converted to a value compatible with $\avi$ with 
$\avi\lb\tjn\rb = \frac{\pi}{2} - \vert \tjn - \frac{\pi}{2}\vert$
\citep[e.g.][]{abbott_gravitational_2017}. In the context of the
inference these are still treated as separate parameters, but we enforce this 
relation in
the Hubble inference prior, which we shall take to be of the following form 
\begin{widetext}
\begin{equation}
	\prio\lb \theta \vert H_0, \lamH\rb 
	= \prio\lb z\vert \lamH\rb \delta\lb \dlum - \dlum\lb z, H_0\rb\rb \delta\lb \avi - \lb\avi\lb \tjn\rb -\tmis\rb\rb  \prio\lb \tjn \vert \lamH\rb,
    \label{eqn:priorModelH0}
\end{equation}
\end{widetext}
where the priors $\prio\lb \dlum \vert H_0,\lamH\rb$ and $\prio\lb \avi\vert \lamH \rb$ are
chosen
to be Dirac delta distributions to account for the different parameters'
correlations. We have already introduced the misalignment angle $\tmis$
as a free parameter in the prior for $\avi$, to account for the possibility
that we have a non-zero shift between $\tjn$ and $\avi$.
In this work, we also
assume that the redshift can be measured with high precision\footnote{Note that there
might be severe peculiar velocity systematics at short distances, such that high precision
might not imply high accuracy, but this is a bias that we are not considering here.}
, to the point that we can
approximate the likelihood/posterior as a Dirac delta distribution.
Inserting the prior, expressing the GW- and EM-data likelihoods in terms of posteriors, 
and integrating out all of the
Dirac delta distributions result in the following reduced form for the combined
data likelihood 
\begin{widetext}
\begin{equation}
	\like\lb D \vert H_0,\lamH\rb
	= \frac{\Psi\lb D\sem, \lambda\sem, D\sgw, \lambda\sgw\rb}{\prio\lb\dlum\lb z_{\text{obs}},H_0\rb \vert\lambda\sem\rb \prio\lb\dlum\lb z_{\text{obs}},H_0\rb \vert\lambda\sgw\rb} \ \text{,} \nonumber
\end{equation}
\begin{equation}
	\text{with} \ \ \Psi
	= \int d\tjn \frac{\prio\lb\tjn\vert\lamH\rb\post\lb \avi\lb\tjn\rb -\tmis,\dlum\lb z_{\text{obs}},H_0\rb\vert D\sem,\lambda\sem\rb \post\lb \tjn,\dlum\lb z_{\text{obs}},H_0\rb\vert D\sgw,\lambda\sgw\rb}{\prio\lb \avi\lb\tjn\rb -\tmis\vert\lambda\sem\rb\prio\lb\tjn\vert\lambda\sgw\rb}	\ \text{.} \label{eqn:dataLike}
    \nonumber
\end{equation}
\end{widetext}
We choose to use a uniform prior for 
the Hubble parameter, $\prio\lb H_0\vert\lamH\rb = \mathcal{U}\lp 0,200\rp$ km s$^{-1}$ Mpc$^{-1}$,
where $\mathcal{U}$ denotes the uniform distribution between two bounds-
The priors for the $\lambda\sgw$- and $\lambda\sem$-models that are used for the posteriors
in the disjoint GW- and EM-analysis are specified in Sec. \ref{sec:gw_mock} and App. \ref{app:em_mock}.

\subsection{Mock data generation}

The posteriors that will be used in the framework described above are
obtained from mock data that is generated with established procedures in
both subsectors and which ensure that we can emulate realistic posteriors
as they would occur in a present-day multi-messenger inference campaign.

\subsubsection{GW sector mock samples} \label{sec:gw_mock}
The mock samples of binary neutron stars are produced considering the masses of the neutron stars as 1.4 $M_\odot$ each, at a luminosity distance $d_L= 21. 56$ Mpc. We have chosen the value of the inclination angle from about zero degrees to 90 degrees for 24 different values of the inclination angle. As we are interested in understanding the systematic error on the Hubble constant (and not the statistical fluctuation), we estimate the impact of different inclination angles for a fixed value of the sky position. We have also chosen the component spin parameters for both the binaries ($\chi_1$ and $\chi_2$) as 0.02. However, the results on the Hubble constant obtained in this analysis are not \gsr{strongly dependent on this choice,} as the spins of the individual objects and the luminosity distance to the source are not degenerate.  For the sky position of the BNS sources, we assume here that the sky position can be inferred from the host galaxy using the information from the electromagnetic counterpart. 

We perform estimation of four parameters namely, the chirp mass $\mathcal{M}$, mass ratio $q$, luminosity distance $d_L$, and inclination angle $\theta_{\rm JN}$ using the publicly available package \texttt{Bilby} \citep{Ashton:2018jfp} with the \mmfix{\texttt{dynesty}} sampler \citep{2020MNRAS.493.3132S} and we assume a three detector configuration (LIGO-Hanford, LIGO-Livingston, and Virgo) \citep{KAGRA:2013rdx, LIGOScientific:2014pky,VIRGO:2014yos, PhysRevD.88.043007} with \mmfix{detector noise consistent with the design sensitivity of LIGO \citep{KAGRA:2013rdx} with a same seed of the noise realization. We have chosen the same seed of the noise realization to demonstrate the systematic errors that can arise for different scenarios of viewing angle mismodeling, and not due to statistical uncertainty for different noise realization.}. The prior on the chirp mass, mass ratio, and distance is taken as uniform $\mathcal{U}[0.4, 4.4] \, M_\odot$, $\mathcal{U}[0.125, 1]$, and $\mathcal{U}[1, 100]$ Mpc respectively. For the inclination angle, we use the
prior $\prio\lb \tjn\vert \lambda\sgw\rb = \frac{1}{2}\Theta\lb \sin\tjn\rb\vert\cos\tjn\vert$ with
the range $[0, \pi]$ rad, \mmfix{where $\Theta(x)$ denotes the Heaviside step function.}

\subsubsection{EM sector mock samples} \label{sec:em_mock}

For the electromagnetic sector, we consider forward shock synchrotron emission from the blast wave of a relativistic GRB jet as the prevailing signal.  This has been the dominant electromagnetic signal from GW170817 since the $\sim2$ week-long kilonova (KN) faded from the infrared bands \citep[e.g.][]{cowperthwaite_electromagnetic_2017}.

We assume a fiducial jet model, compute its emission with the \afterglowpy{} {\tt v0.8.0} software package \citep[][]{ryan_gamma-ray_2020}, and generate mock radio, optical, and X-ray observations for an idealized observing scenario. We also include mock measurements of the flux centroid position in our mock radio observations, since they contain key information regarding the jet inclination. To focus on systematics and biases inherited from the jet modeling, we ignore KN emission and many realistic confounding factors such as availability of facilities, sun constraint, excess extinction, and source confusion. The approach is similar to the one used in \cite{gianfagna_joint_2023,gianfagna_potential_2024}.

\gsrfix{GRB afterglows are produced by synchrotron emission from the forward shock of the GRB blast wave as it impacts the surrounding medium.  During the afterglow phase, the jet evolution is determined by the distribution of energy in the jet, which is typically assumed to be axisymmetric and described by a function $E(\theta)$ in the angle from the jet axis $\theta$.}
Depending on the particular jet structure, $E(\theta)$ may have several parameters. In this work we use a Gaussian jet (GJ) $E = E_0 \exp(-\theta^2/2\theta_c^2)$ with central energy $E_0$, core width $\theta_c$, and truncated at an outer angle $\theta_w$. We also use a power-law jet (PLJ) $E = E_0 (1+\theta^2/b\theta_c^2)^{-b/2}$ that utilizes the same $E_0$, $\theta_c$, and $\theta_w$ plus the power-law index $b$.
The energy distribution is the key difference between the considered jet models.  \gsrfix{To fully specify an afterglow model we must also set an ambient external density $n_0$, and four parameters for the synchrotron radiation ($p$, $\epsilon_e$, $\epsilon_B$, and $\xi_N$, see \citep[e.g.][]{granot_shape_2002}). 
\gsrfix{To form observations the system is assigned a redshift $z$, luminosity distance $\dlum$, and the jet is inclined $\avi$ relative to the line of sight.} 
Specifically for mock centroid position data, the system is also given initial sky coordinates and a position angle (${\rm RA}_0$, ${\rm Dec}_0$, ${\rm PA}$).  For further details on the afterglow model see Appendix \ref{app:em_mock}.}

A Gaussian jet is then described by 14 parameters \gsrfix{($\avi$, $z$, $\dlum$, $E_0$, $\theta_c$, $\theta_w$, $n_0$, $p$, $\epsilon_e$, $\epsilon_B$, $\xi_N$, ${\rm RA}_0$, ${\rm Dec}_0$, ${\rm PA}$)} and a power-law jet additionally includes the parameter $b$.  For the Hubble inference, all but $\{z, \avi, d_L\}$ are nuisance parameters\gsrfix{, and $z$ is fixed to $z^{(\text{true})}$.  We leave all parameters free (except $z$) during the EM parameter estimation on the mock data, and marginalize over the nuisance parameters for $H_0$ analysis.}

We use a single fixed jet model to generate our mock data, the fiducial Gaussian jet model of \citet{ryan_modelling_2023}, which is a good fit to the full set of GW170817 afterglow observations.
Once the jet model is specified, we generate mock data by simulating an idealized observation campaign using radio, optical, and x-ray photometric observations as well as radio observations of the centroid position. The data generation algorithm is designed to model an optimistic observation scenario for an EM counterpart and details on the algorithm are provided in App. \ref{app:em_mock}, together with the 14 parameters of the injected GJ.

We obtain $9$ EM events with an inherent Gaussian jet structure and viewing
angles ranging from $0\degree-80\degree$. These datasets are then fit with either a GJ or
PLJ model and we consider two inference scenarios, one where only the lightcurve data is
fit, and another where mock \mm{radio} data for the centroid motion is included, resulting in
a total of $4\times9$ posteriors. The used priors are provided in App. \ref{app:em_mock}.
We label the posterior sets with the structure model that
was employed for the fit and add the post-fix ``c'' for posteriors that include input from
the centroid motion: \texttt{GJ}/\texttt{PLJ} and \texttt{GJc}/\texttt{PLJc}. 

\section{Results and discussion} \label{sec:results}

Based on the mock data generated with the procedures outlined in Sec. \ref{sec:gw_mock}
and \ref{sec:em_mock} (see also Fig. \ref{fig:mock-lc} for the mock light curves), 
we conduct several inference campaigns to probe the \textit{model}
and \textit{misalignment} bias respectively. For reference, we compute $H_0$ posteriors 
only using the GW data and the knowledge about the redshift, and the results will be
labeled with \texttt{GW+z}. On the EM side, we consider four different scenarios, which either
use a GJ (as in the injection) or a PLJ model for the fitting, and for
each jet structure we compute posteriors once only from the lightcurve fit 
(denoted \texttt{GJ}/\texttt{PLJ})
and then also
with the inclusion of centroid data (denoted \texttt{GJc}/\texttt{PLJc}). Once we combine the EM posteriors with the GW and
redshift information, we will label the four different scenarios with
\texttt{GW+GJ+z} and \texttt{GW+PLJ+z} for the lightcurve-only fits and
\texttt{GW+GJc+z} and \texttt{GW+PLJc+z} for the posteriors including the centroid motion.
Since we are only working with system inclinations $<90\degree$, the definitions of
$\tjn$ and $\avi$ agree with each other and we shall quote all results in terms of the
jet viewing angle $\avi$ (and the misalignment angle $\tmis$ if it is non-zero). Below we describe the mismatch that can arise due to error in modeling the jet and intrinsic mismatch between the inclination angle and jet angle in Sec. \ref{sec:model-bias} and Sec. \ref{sec:misamatch-bias} respectively.

\subsection{Aligned runs and afterglow model bias}  \label{sec:model-bias}

\gsrfix{For the analysis of the \textit{model} bias, we combine GW and EM data without using a misalignment angle, i.e. $\tmis^{(\text{true})} = 0$.
The median and $2.5$th-$97.5$th percentiles for these posteriors, as well as relevant statistics are tabulated in Tab. \ref{tab:no-offset-H0-HPD} and \ref{tab:no-offset-H0-HPD-stats} in App. \ref{app:data-supp}.}

\gsrfix{The individual reference runs without EM observations, \texttt{GW+z}, demonstrate a $\mathcal{O}(1-2\sigma)$ difference in $H_0$ inference at small viewing angles which decreases to $\lesssim \mathcal{O}(0.5\sigma)$ as $\tjn$ increases past $40^\circ$. This can be attributed to the $\dlum$-$\tjn$-anti-correlation, which is strongest at smaller viewing angles, and the noise in the mock data, which can easily shift the posterior peak along the $\dlum$-$\tjn$ degeneracy.  Since our mock GW data uses a single noise realization, all \texttt{GW+z} runs inherit an identical push to larger $\tjn$ (smaller $\dlum$) and hence larger $H_0$.
As the viewing angle increases past $\gtrsim 50^\circ$ the two GW polarizations decouple in intensity, which breaks the $\dlum$-$\tjn$-degeneracy and increases the $H_0$ precision to $\mathcal{O}(\leq 10\%)$ \citep{chen_viewing_2019,nakar_afterglow_2021}.} 

\gsrfix{The effect of EM afterglow observations on $H_0$ inference depends on how well the EM sector constrains $\avi$ relative to the GW data \smn{(see the results in Tab. \ref{tab:no-offset-H0-HPD} and \ref{tab:no-offset-H0-HPD-stats} in App. \ref{app:data-supp}.}).  At small inclinations, $\avi \leq 20^\circ$, the EM data is bright and highly constraining.  All runs show a median $H_0$ within $\sim 0.5\sigma$ of the true $H_0$ with $\sigma / H_0 \lesssim 3\%$, substantial improvement over \texttt{GW+z}.  The photometric-only EM runs (\texttt{GW+GJ+z} and \texttt{GW+PLJ+z}) perform \textit{worse} than \texttt{GW+z} for $\avi \geq 30^\circ$, constraining $H_0$ to only $\sim 10\%$ with differences as large as $1.5 \sigma$.  Runs including centroid position observations (\texttt{GW+GJc+z} and \texttt{GW+PLJc+z}) remain constraining for $\avi = 30^\circ, 40^\circ$, with similar difference ($\mathcal{O}(0.5\sigma)$) but improved precision ($\sim$few \%).   As $\avi$ increases past $50^\circ$, even at our very close fiducial luminosity distance, the afterglow emission becomes too faint to observe and the mock datasets become almost entirely upper limits, see Fig. \ref{fig:mock-lc}. At this point the only added information from the EM posteriors can be an improved lower bound on the inclination, which translates into a lower bound on $H_0$.} 

\begin{figure*}[!ht]
    \begin{center}
        \setlength\extrarowheight{-3pt}
        \begin{tabular}{cc}
             \includegraphics[trim=0 0 1.5cm 11.5cm, clip,scale=0.38]{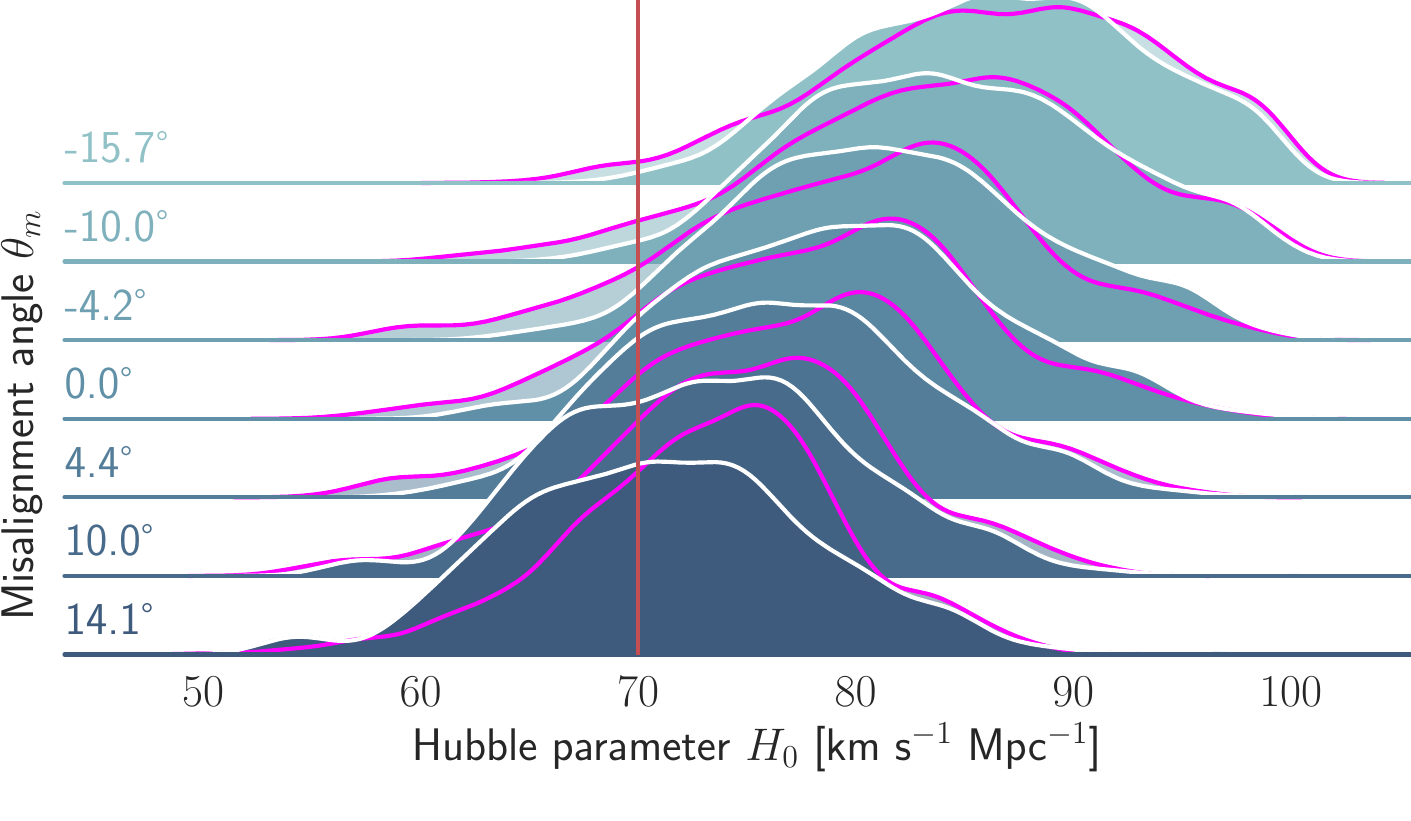} &  
             \includegraphics[trim=1.18cm 0 2.0cm 11.5cm, clip,scale=0.38]{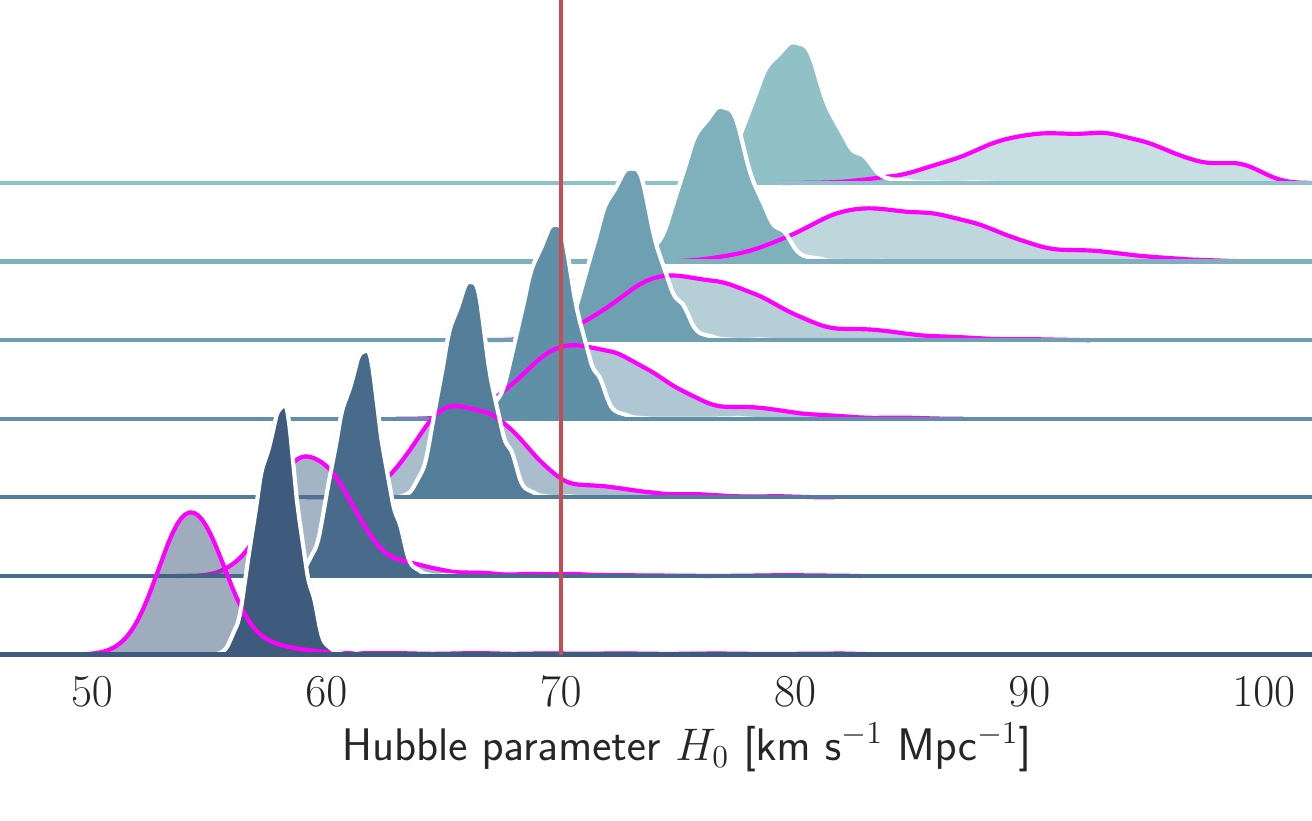}
        \end{tabular}
    \end{center}
\caption{
    Progression of joint \texttt{GW+GJ(c)+z} Hubble posteriors for $\tjn^{\text{(true)}} = \avi^{(true)} + \tmis$,
    with $\tmis$ varying systematically from $-16\degree$ to $14\degree$. We show the posteriors for $\avi^{(true)}=30\degree$ (white contours) and  $\avi^{(true)}=40\degree$ (magenta contours). The posteriors on the left and right represent the results from the inference with the \texttt{GJ} and \texttt{GJc} scenario respectively.}
\label{fig:offset-G4EM3-EM9}
\end{figure*}

\gsrfix{Fig. \ref{fig:offset-G4EM3-EM9} (the $\theta_m=0$ row) shows the $H_0$ posteriors for $\avi = 30^\circ, 40^\circ$, including the \smn{less (and more) precise posterior on $H_0$ for the \texttt{GJ} (and \texttt{GJc}) case}. A similar behavior is also identified in the analysis of the GW170817 afterglow in \cite{gianfagna_potential_2024}, where it is attributed to an attempt of the jet models to accommodate a late-time excess in the light curve. However, since we also find this for the fits of our mock observations which do not include such an effect, the behaviour may be more generic, possibly due to parameter degeneracies altering the available prior volume (see \citet{ryan_modelling_2023} for discussion). } 

\gsrfix{To assess the presence of model bias we compare the $H_0$ posterior distributions from fits run on the same dataset with different assumed jet models.  Table \ref{tab:no-offset-H0-HPD-stats} contains the difference in median $H_0$ values from \texttt{GJ} and \texttt{PLJ} runs normalized by the \texttt{GJ} $H_0$ ($\Delta H_0$) and by the width $\sigma$ of the \texttt{GJ} $H_0$ posterior ($\delta H_0$).  In the no-centroid runs, \texttt{GJ} and \texttt{PLJ} models show a maximum bias of 2\% in median $H_0$.  This occurs in the $\avi = 20^\circ$ run, and corresponds to a $0.49\sigma$ bias.  All other runs show less than $0.08\sigma$ bias. }
\gsrfix{Including astrometric centroid data increases the constraining power of the EM afterglow, increasing the precision of $H_0$ inference, which also increases its vulnerability to bias from model mis-specification.  Between the \texttt{GW+GJc+z} and \texttt{GW+PLJ+z} runs we find again a maximum discrepancy in median $H_0$ of 2\%, this time in the $\avi=40^\circ$ run.} 

\subsection{Angular mismatch bias} \label{sec:misamatch-bias}

\begin{figure*}[ht]
    \begin{center}
        \includegraphics[trim=1.5cm 0 0 0,clip,scale=0.61]{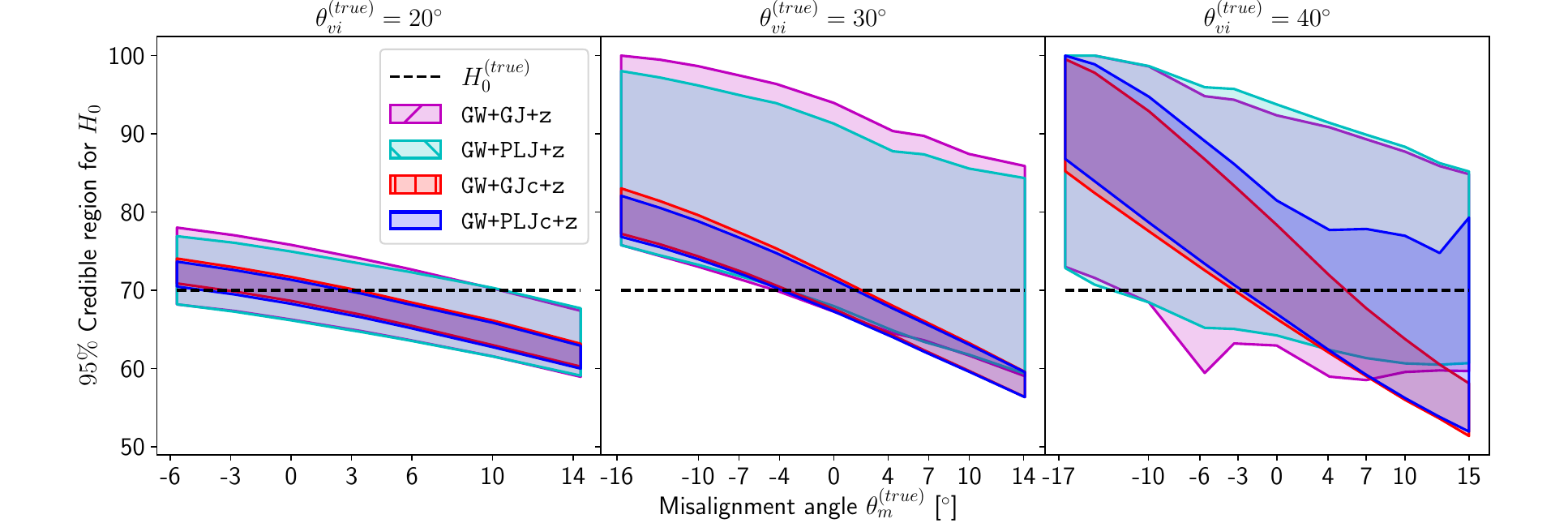}
    \end{center}
    \caption{
        $95\%$ HPD credible regions ($2\sigma$) for the
         Hubble posteriors of all 4 GW+EM inference scenarios (different scenarios are
         indicated by red/magenta colored regions for the GJ(c) model and blue/cyan for
         the PLJ(c) model), with $\avi^{\text{(true)}}=20\degree,30\degree,40\degree$ 
         and $\tjn^{\text{(true)}} = \avi^{\text{(true)}} + \tmis$ and $\tmis$ varying 
         systematically. The truth is indicated with a dashed line. (data tabulated in
         Tab. \ref{tab:offsetGrid} and \ref{tab:offsetGrid-stats})
    }
    \label{fig:offsetGrid}
\end{figure*}

We now turn our attention to systems that have a non-vanishing line-of-sight misalignment angle, and compute
joint GW+EM posteriors for the Hubble parameter neglecting this fact, i.e. we will still use
$\tmis = 0$ in Eq. \eqref{eqn:priorModelH0}. The mock data in the GW sector was set up with this
in mind and we computed several slightly shifted GW posteriors, scattered around the events
that were used in Sec. \ref{sec:model-bias}. Consequently, $\avi^{(\text{true})}$ will be fixed
and events with deviating $\tjn^{(\text{true})}$ are combined in the joint posterior, resulting
in a systematic variation of $\tmis^{(\text{true})}$. As discussed in Sec. \ref{sec:intro},
increasing $\tmis^{(\text{true})}$ implies that the GW posterior is shifted toward subsequently
higher inclination regions and the overlap with the EM posterior will occur at larger values of
$\dlum$, thus we expect the $H_0$ posterior to shift toward smaller $H_0$ for greater misalignment
$\tmis^{(\text{true})}$, this behavior is demonstrated in Fig. \ref{fig:offset-G4EM3-EM9}, for
the posteriors from the \texttt{GW+GJ(c)+z} scenarios at $\avi^{(\text{true})}=30\degree$ and $40\degree$. Two
important observations can be made: Clearly, the high-precision
posteriors from the \texttt{GJc} analysis ($\mathcal{O}(1-2\%)$) are more sensitive to the misalignment and
already a small shift of $\pm 4\degree$ causes a $2-4\sigma$ bias in the resulting $H_0$ (cf. Tab. \ref{tab:offsetGrid-stats}).
Furthermore, we know from Sec. \ref{sec:model-bias} that the posterior for the \texttt{GJ} analysis
is biased towards larger $H_0$, which is evident in Fig. \ref{fig:offset-G4EM3-EM9}, and we
note that a large and positive $\tmis^{(\text{true})}$ will result in a more accurate
Hubble posterior, since the two effects cancel. On the other hand, a small negative 
$\tmis^{(\text{true})}$ can amplify the model bias, leading to an even more inaccurate inference.
We also confirm this behavior through the change of the $95\%$ credible regions for the cases 
$\avi^{(\text{true})}=20\degree$ and $40\degree$,
see Fig. \ref{fig:offsetGrid},
which seem to suggest the general trend that for an inference that is using lightcurve and
centroid data $\vert\tmis^{(\text{true})}\vert \sim 3\degree-6\degree$
will lead to a $\mathcal{O}(1-2\sigma)$ bias, which is then quickly increasing for larger magnitudes\footnote{
    Note that the median and $2.5$th-$97.5$th percentiles for the posteriors in Fig. \ref{fig:offsetGrid}
    are tabulated in Tab. \ref{tab:offsetGrid} and useful statistics for the Hubble posterior
    are listed in Tab. \ref{tab:offsetGrid-stats}.
}.
For a viewing angle of $20\degree$ we can also state that the inference using only photometry
data is less sensitive since $\vert\tmis^{(\text{true})}\vert \leq 6\degree$ will not bias the
result beyond $\mathcal{O}(1\sigma)$, however, we cannot make a similar result for the higher-inclination
systems shown in Fig. \ref{fig:offset-G4EM3-EM9} and \ref{fig:offsetGrid}, since they have
an intrinsic bias that can combine with the misalignment to cause more drastic shifts.

\begin{figure*}[ht]
    \begin{center}
        \includegraphics[trim=1cm 0.72cm 1cm 0,clip,scale=0.58]{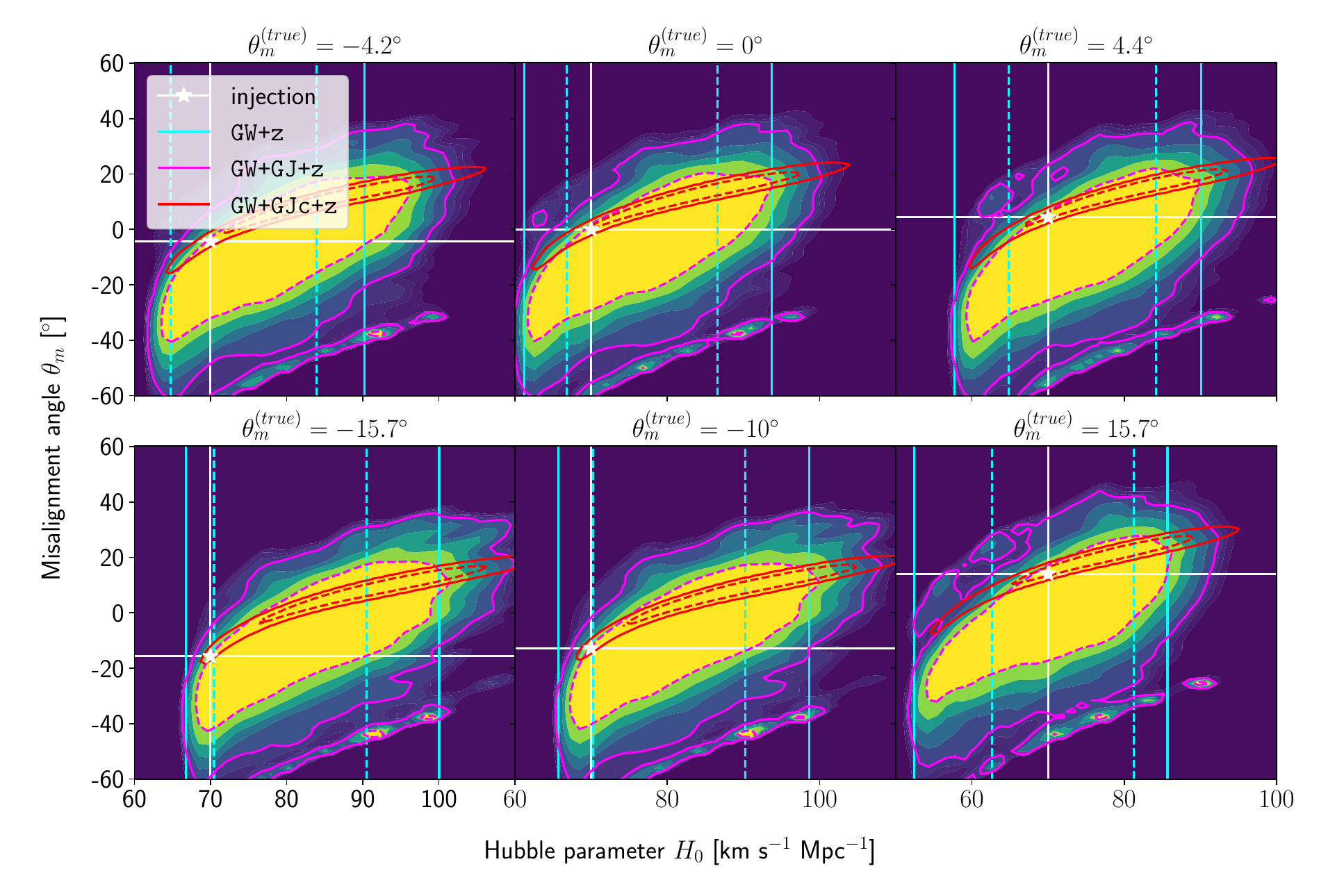}
    \end{center}
    \caption{
        $2$D posteriors for the $(\tmis,H_0)$-inference of the $\avi^{(\text{true})}=30\degree$
        event, with \mmfix{$\tmis^{(\text{true})} \in\lc -15.7\degree,-10\degree,-4.2\degree,0\degree,4.4\degree,15.7\degree\rc$} from left to right. The
        heat map represents the \texttt{GW+GJ+z} posterior, while the magenta and red contours indicate
        the $1\sigma$ (dashed) and $2\sigma$ (solid) HPD regions for the \texttt{GW+GJ+z} and \texttt{GW+GJc+z}
        posteriors respectively. The truths are marked with white stars. We also indicate the $1\sigma$ (dashed) and
        $2\sigma$ (solid) regions of the \texttt{GW+z} posteriors with cyan lines.
    }
    \label{fig:mitigationGrid}
\end{figure*}

\subsection{Mitigation of angular mismatch bias} \label{sec:mitigation}

Now that we have established that even a small \gsr{line-of-sight} angular mismatch of 
$\vert\tmis^{(\text{true})}\vert \sim 3\degree$
can lead to a severely biased Hubble posterior, we want to address a possible method of
mitigating this bias. Since we have a well-defined mismatch parameter $\tmis$, we can
consider this as a nuisance parameter in the joint inference and co-infer it with the Hubble
parameter. Marginalization of the posterior in the $H_0$-$\tmis$-plane should then lead
to an unbiased posterior for the Hubble parameter and we can also obtain a posterior for
the mismatch. The key modification
is that we treat $\tmis$ as a variable with an associated prior added to
Eq. \eqref{eqn:priorModelH0}.
The effect of the misalignment angle is to shift
the $\dlum$-$\avi$-posterior/-likelihood from the EM inference in the $\avi$-direction,
without changing its shape. Consequently, in the joint posterior, we will compute a
shifted overlap in $\tjn$-space in Eq. \eqref{eqn:dataLike}.
Note that contrary to the behavior of $\tmis^{(\text{true})}$ in Sec. \ref{sec:misamatch-bias},
$\tmis$ shifts the EM likelihood with respect to the \texttt{GW+z} likelihood, for $\tmis > 0$ it
is shifted to larger $\avi$ leading to a posterior overlap with the GW posterior that
occurs at larger inclination than for $\tmis=0$ and therefore at smaller $\dlum$,
which will increase the inferred $H_0$, thus we expect a direct positive correlation between
$\tmis$ and $H_0$ (see also Fig. \ref{fig:tmisEffect} for a qualitative comparison).

We apply the modified inference
to the same cases at $\avi^{(\text{true})}=30\degree$, as shown in Fig. \ref{fig:offset-G4EM3-EM9} and \ref{fig:offsetGrid}, 
using the prior
\mmfix{$\prio\lb\tmis\vert\lamH\rb = \mathcal{U}\lp -60\degree,60\degree\rp$} rad for the misalignment angle and
considering only the Gaussian jet model with or without centroid. The resulting $2$D
posteriors in the $\tmis-H_0$-plane for the cases of \mmfix{$\tmis^{(\text{true})}\sim0\degree,\pm 4\degree,-10\degree,\pm 16\degree$}
are shown in Fig. \ref{fig:mitigationGrid}. Considering the case of $\tmis^{(\text{true})}=0\degree$ first, we observe the expected positive
correlation/degeneracy, and both the \texttt{GW+GJ+z} and \texttt{GW+GJc+z} inference are consistent with 
the truth to within $1\sigma$ respectively. 

For the scenarios with intrinsic misalignment, it seems that
negative misalignment affects the accuracy worse than positive misalignment, while all posteriors from the
\texttt{GJ} datasets lead to a $\mathcal{O}(1\sigma)$ accuracy, the large negative misalignments  $\tmis^{(\text{true})} \leq -10\degree$
are at the edge. This behavior is more pronounced in the inference with the \texttt{GJc} datasets, where all scenarios
with $\tmis^{(\text{true})} \leq -4\degree$ are only accurate to within $2\sigma$, while positive
misalignment still maintains a $\mathcal{O}(1\sigma)$ accuracy \mmfix{(see Fig. \ref{fig:marginal-mitigationGrid} and \ref{fig:comparison-stats-inference-modes} 
for statistics of the marginal $H_0$ posteriors)}. \mmfix{This is related to our choice of using the same
noise seed for all GW mock data realizations\footnote{This is chosen in order to remove the statistical fluctuation due to variation in the noise seed and demonstrate the systematic bias due to EM modeling misalignment}, which for the considered data shifts the $\dlum$-$\avi$ posterior
towards larger inclination angles than the injected value\footnote{Note that the effect of $\tmis$ in Eq. \eqref{eqn:priorModelH0} is to shift the likelihood in the opposite direction than what is shown in Fig. \ref{fig:tmisEffect}, thus, $\tmis>0$
implies a shift of the EM posterior towards larger $\avi$.
}, 
see the discussion in Sec. \ref{sec:model-bias}. Consequently,
the combined GW$+$EM posteriors tend to have higher posterior overlap for positive misalignment angles, thus leading
to more posterior support towards $\tmis>0$.}
However, all posteriors
have support over a large region in the $\tmis-H_0$-plane and once we compute the marginal
posteriors for the Hubble parameter, we shall obtain unconstraining posteriors, due
to the degeneracy of $\tmis$ and $H_0$.  \mmfix{The statistics for the marginal $H_0$ posteriors
in Fig. \ref{fig:mitigationGrid} demonstrate that the precision,
even for the most constraining EM datasets (\texttt{GJc}), is reduced to within a factor of
$1.1-1.2$ of the level of the
\texttt{GW+z} posteriors (see column 2 in Fig. \ref{fig:comparison-stats-inference-modes}) 
}

Considering the most precise posteriors that are obtained from analyzing lightcurve and centroid
data, we obtain $\mathcal{O}(\geq 10\%)$ constraints for $H_0$ from marginalizing $\tmis$ in the $2$D posteriors
in Fig. \ref{fig:mitigationGrid}, which is comparable to the constraints
obtained by only fitting the lightcurve in the cases with zero misalignment in Sec \ref{sec:model-bias}. 
We also compute $2$D posteriors in the $\tmis-H_0$-plane for the series of events without
misalignment that was considered in Sec. \ref{sec:model-bias}, where we would expect to obtain
an accurate inference of $H_0$ with the maximum posterior region centered around $\tmis = 0$.
This is indeed what we find, $\tmis$ is constrained to zero with $\mathcal{O}(1-2\sigma)$ precision and
we obtain Hubble posteriors that are accurate to within $1.5\sigma$ for all viewing angles 
$\avi^{(\text{true})}=3\degree-80\degree$, the posterior statistics are \mmfix{summarized in
Tab. \ref{tab:tmisZeroH0}  and column 2 of Fig. \ref{fig:tmisZero-comparison-stats-inference-modes}
in App. \ref{app:data-supp}. However,
also for these scenarios the gain in precision from including EM data is reduced due to marginalization
of the misalignment angle, see column 2 of Fig. \ref{fig:tmisZero-comparison-stats-inference-modes}
for more details.}

Trusting on the theoretical models, we might expect
that the physical misalignment will be of $\mathcal{O}(\text{few}\degree)$ at most, consequently,
a narrower choice for the $\tmis$-prior could be motivated, e.g. 
$\prio\lb\tmis\vert\lamH\rb = \mathcal{U}\lp -10\degree,10\degree\rp$, and this could
also reduce the spread in the posterior, while still being able to confidently infer
the misalignment in the cases with $\vert\tmis^{(\text{true})}\vert \sim 3\degree-7\degree$. 
\mmfix{We have applied the same bias-mitigating procedure as discussed above to datasets with
$\tmis^{(\text{true})} \in\lc -6.2\degree,-4.2\degree,0\degree,4.4\degree,6.2\degree\rc$,
but we applied the narrower prior range for $\tmis$ and the statistics for the marginal $H_0$
posterior indeed show that for the posteriors from the \texttt{GJc} datasets the accuracy behaves similar
to the case with a wider $\tmis$ prior, but the precision is again improved to a factor of two
compared to the \texttt{GW+z} posteriors, Fig. \ref{fig:marginal-mitigationGrid-narrow-tmis}, 
as well as column 3 in Fig. \ref{fig:comparison-stats-inference-modes}.} 
\mmfix{This demonstrates the need for narrow theoretical bounds for the misalignment angle in the joint
inference of $H_0$, for the amount of precision that can be achieved in the single-event level inference of $H_0$
in this scenario is directly linked to the prior width in $H_0$.}

\section{Conclusion and Future Outlook}

\mm{
In this work, we analyze scenarios for the joint inference of the Hubble parameter from GW and EM data at the single-event level that attempt to mimic a realistic multi-messenger campaign in the near future, in terms of available data and models, under ideal conditions.
}
We point out a potential source of bias in the inference of the value of $H_0$ from multi-messenger observation due to incorrect model fitting with EM observations and/or intrinsic misalignment between the inclination angle and the jet angle. For the case of incorrect modeling, we find that over the range $\avi=3\degree-40\degree$ the \textit{model} bias is a subdominant effect as both assumptions
on the jet structure lead to consistent results to within $\mathcal{O}(0.5\sigma)$, extending the findings
for a GW170817-like event in \cite{gianfagna_potential_2024}. \texttt{GJc/PLJc} are precise and accurate
for $\avi=3\degree-40\degree$, while \texttt{GJ/PLJ} are precise and accurate for $\avi=3\degree-20\degree$
(although already by a factor of $3$ worse than the analysis with centroid data), but less
precise ($\mathcal{O}(13\%-20\%)$) and inaccurate at $\avi=30\degree-40\degree$, with an $\mathcal{O}(1.2\sigma) = \mathcal{O}(10\%)$
overestimation, \mm{which can be prohibitive for cosmological inference.}

On the other hand, the physical misalignment between the orbital angular momentum axis of the inspiraling binary
and the jet axis was identified as a significant source of bias in the joint inference of GW and
EM data, and particularly, when centroid information is included. Data with line-of-sight misalignment $\vert\tmis^{(\text{true})}\vert \sim 3\degree-7\degree$ 
that is analyzed under the assumption of alignment, which
is frequently employed (see \cite{abbott_gravitational_2017,chen_viewing_2019,farah_counting_2020})
as the misalignment is not constrained by current theoretical models, can lead to 
$\mathcal{O}(1-2\sigma)$ biases in $H_0$ corresponding to a $3-7\%$ deviations for $\avi^{(\text{true})}=20\degree-30\degree$
and $5-12\%$ for $\avi^{(\text{true})}=40\degree$.
This shows that presently used models for the EM counterpart lead to posteriors that are even
more sensitive than what was found by \cite{chen_systematic_2020}, based on Gaussian estimates.
Since we are using super-ideal events, this is
a lower limit and it is already close to the required accuracy for resolving the Hubble tension
at about $\gtrsim 3\sigma$ \citep[e.g.][]{hotokezaka_hubble_2018}. Ensuring an accurate $H_0$ inference from bright sirens would therefore require a
method to mitigate such systematic biases. 

Note that random values of misalignment for a large number of \gsr{events} would also cause random
systematic errors in $H_0$, and if multiple events were combined their effects \gsr{may average out
to provide an unbiased $H_0$. However, }the prospects for observing many BNS or BHNS mergers that
can be analyzed with the \textit{bright sirens} method are limited and the expectation is that
analyses in the near future will have access to a few events at best, see e.g.
\cite{gianfagna_potential_2024}.  \mm{
In this case, even a few observations with precise but inaccurate $H_0$-posteriors, from the inclusion
light curve and centroid data, could dominate over other bright siren measurements of $H_0$ that only have access to 
the redshift and the bias of the multi-event inference would be driven by the systematic errors in
the analysis at the single-event level. Consequently, it is paramount to identify biases at the single-event level and to reduce these if possible.
}

To mitigate the uncertainty due to the \textit{model} or \textit{misalignment} biases, one is required to jointly infer the value of $\theta_m$ and $H_0$ to obtain an unbiased marginal posterior on $H_0$, and also an inference of $\theta_m$. The multi-messenger observation from GW and EM data, is capable to provide information on $\theta_m$, and our analysis shows that it should be inferred together, rather than assuming the value of $\theta_m$ to be zero. The later assumption can cause a biased $H_0$ even for a minor deviation that is easily feasible astrophysically \citep[e.g.][]{stone_pulsations_2013,li__quasiperiodic_2023,hayashi_general-relativistic_2022,nagakura_jet_2014}.  

We conclude that co-inferring and marginalizing over the misalignment angle can indeed reduce
the bias in the inferred $H_0$ to within $2\sigma$, however at the cost of the precision that
is available at the single-event level.
\mmfix{If the misalignment angle is given a wide prior range this can reduce the precision of
the GW+EM inference to the level of the GW-only inference, thus removing the benefit of the combined analysis for the considered  scenarios.  However, this might change at larger redshifts, depending on the
quality of the EM data compared to the GW data. Reducing the prior range for $\tmis$ can restore
the gain in precision for the combined GW+EM inference up to a factor of $2$ with respect to the
\texttt{GW+z} posteriors, thus improvements on the theoretical bracket for this parameter will be
paramount for ensuring a highly constraining and unbiased inference.
}

In the future, it will be important to incorporate this analysis in the bright siren studies for which EM measurements of the viewing angle from light-curve and/or centroid data is available. Along with the inference of the $H_0$, this technique makes it possible to also infer the value of the mismatch angle $\theta_m$. 
This can be useful in understanding the astrophysical properties of BNS/BHNS mergers better, shedding light on to what extent merger-driven jets are aligned with the orbital angular momentum. 
With a population of BNS/BHNS events, a population-level inference of $\theta_m$ can lead to new insights about the complexity of the post-merger phase.

\textbf{Acknowledgement} The authors thank Samsuzzaman Afroz for useful comments on the draft. This research was supported in part by Perimeter Institute for Theoretical Physics. Research at Perimeter Institute is supported in part by the Government of Canada through the Department of Innovation, Science and Economic Development and by the Province of Ontario through the Ministry of Colleges and Universities. M.M. acknowledges the support of NSERC, funding reference number No. RGPIN-2019-04684. The work of S.M. is part of the \texttt{⟨data|theory⟩ Universe-Lab}, supported by TIFR and the Department of Atomic Energy, Government of India. The authors express gratitude to the computer cluster of \texttt{⟨data|theory⟩ Universe-Lab}. The authors would like to thank the  LIGO/Virgo/KAGRA scientific collaboration for providing the data. LIGO is funded by the U.S. National Science Foundation. Virgo is funded by the French Centre National de Recherche Scientifique (CNRS), the Italian Istituto Nazionale della Fisica Nucleare (INFN), and the Dutch Nikhef, with contributions by Polish and Hungarian institutes. This material is based upon work supported by NSF's LIGO Laboratory which is a major facility fully funded by the National Science Foundation.
\pagebreak{}
\appendix

\section{EM mock data generation} \label{app:em_mock}

\gsrfix{GRB afterglows are produced by synchrotron emission from a non-thermal electron population, accelerated at the forward shock driven by the GRB jet.  
The jet launching mechanism imbues the jet with a distribution of energy $E(\theta)$, typically taken to be axisymmetric.  The jet sweeps up the ambient medium of density $n_0$, forms a shock, decelerates, and spreads laterally.  
As first-principles simulations of BNS mergers have not converged on a jet structure, we instead use simple phenomenological models: a Gaussian jet (GJ) $E = E_0 \exp(-\theta^2/2\theta_c^2)$ and a power-law jet (PLJ) $E = E_0 (1+\theta^2/b\theta_c^2)^{-b/2}$.  Both have central energy $E_0$, core width $\theta_c$, and are truncated at an outer angle $\theta_w$, the PLJ also includes an index $b$.} 

\gsrfix{Following standard afterglow theory \citep[e.g.][]{granot_shape_2002}, the non-thermal electrons acquire a power-law energy distribution of index $-p$, made up of a fraction $\xi_N$ of the available electrons.  The non-thermal electrons and the magnetic field are each assigned fractions $\epsilon_e$ and $\epsilon_B$, respectively, of the available shock energy.  The GRB is placed at redshift $z$ and luminosity distance $d_L$ and oriented $\avi$ relative to Earth.  For pure photometric data this is enough to compute a flux, for astrometric centroid data we also need initial sky coordinates RA$_0$ and Dec$_0$ and a position angle PA on the sky.}

We use the fiducial Gaussian jet model of \citet{ryan_modelling_2023} for the generation of light curves. This model
is consistent with the GW170817 afterglow observations and has the parameters $E_0=4.8 \times 10^{53}$ erg, $\theta_c = 3.2^\circ$, $\theta_w=22.4^\circ$, $n_0=2.4\times10^{-3}$ cm$^{-3}$, $p=2.13$, $\epsilon_e=1.9\times10^{-3}$, $\epsilon_B=5.8\times10^{-4}$, and $\xi_N=0.95$. Emission from this jet model is computed using \afterglowpy{} {\tt v0.8.0} with jet spreading enabled, see \citet{ryan_gamma-ray_2020} and \citet{ryan_modelling_2023} for implementation details.

We generate mock data from the fluxes of the injected GJ, by simulating an idealized observation campaign using radio, optical, and x-ray photometric observations as well as radio observations of the centroid position.  We designed the data generation algorithm to roughly model an optimistic observation scenario for a GW electromagnetic counterpart. 

    The procedure goes as follows: for each photometric band, we space observations geometrically between 1 and $10^4$ days post-merger.  Starting at $t$ = 1 day we compute the model flux, add noise, and determine whether our mock observatory would detect the emission.  If the source is undetected an upper limit is recorded and the next observation is attempted a factor $10^{1/2} \approx 3.1 $ later in time, giving these monitoring observations a cadence of two per decade in time.  If at any time the source is detected, a flux observation with relevant uncertainty is recorded, and the next observation is scheduled a factor $10^{1/4} \approx 1.8$ later in time, giving an observing cadence of four per decade in time.  Once the afterglow is detected in a band, ``observations'' continue in that band at the four per decade cadence until two successive upper limits are found, at which point observations in that band cease.  Radio VLBI observations are carried out between 10 and $10^4$ days post-merger at the two observations per decade cadence.  This approach generates sufficiently dense observations to measure the initial rising slope of the light curve (if the emission is sufficiently bright) and covers a duration of time over which the jet break is expected to appear.  These two features contain the most geometric information in the light curve, and if detectable are essential for extracting the inclination from GRB afterglow observations.

Our mock radio, optical, and X-ray observatories emulate optimistic observations from VLA, HST, and Chandra, respectively, with uncertainties and noise levels similar to what was obtained with GW170817 \citep[e.g.][]{fong_optical_2019,troja_accurate_2021}.  We characterize the radio and optical observatories with a flux limit $F_{\rm lim}$ (representing the noise level of the instrument) and a fractional source uncertainty $\delta$ (representing calibration and background subtraction uncertainties). We assign a total uncertainty to each observation by simply adding these in quadrature: $\sigma = \sqrt{F_{\rm lim}^2 + \delta^2 F_{\rm src}^2}$, where $F_{\rm src}$ is the model afterglow flux.  The synthetic observed flux $F_{\rm obs}$ is $F_{\rm src}$ plus a random noise contribution drawn from a normal distribution of mean 0 and width $\sigma$.   If $F_{\rm obs} < 3 F_{\rm lim}$ the observation is declared a non-detection and we emit an upper limit of $3 F_{\rm lim}$.  Otherwise the observation is a detection and we emit $F_{\rm obs}$ with the uncertainty $\sigma$.  In parameter estimation these observations are assigned a Gaussian likelihood.  Our radio photometry is generated at $\nu =$ 3 GHz with $\delta = 0.05$ and $F_{\rm lim} = 3$ $\mu$Jy, representing a long integration by a VLA-like instrument.  Our optical photometry is generated at $\nu = 5.1 \times 10^{14}$ Hz with $\delta = 0.05$ and $F_{\rm lim} = 5$ nJy, representing a long integration by a HST-like instrument in the F606W filter. 

Since many of these sources are faint, we generate mock X-ray data with Poisson statistics unless we exceed 100 observed counts. For each observation we compute $F_{\rm src}$ at 5 keV and convert this to an expected number of counts $S$ by multiplying by a flux-to-counts conversion factor of $1.4\times10^8$ mJy$^{-1}$, a typical value for a 100 ks Chandra observation of GW170817 \citep{troja_accurate_2021}.  We then draw a random background count $B$ from a normal distribution of mean $1.0$ and width $0.5$, truncated to $B > 0$ (again, typical for GW170817).  The observed photon count $N$ is then drawn from a Poisson distribution with mean $B+S$.  If $N > 100$ we replace this Poisson observation with a Normally distributed flux observation of width $0.1 \times F_{\rm src}$.  In parameter estimation X-ray observations are assigned a Poisson likelihood following the approach of \citet{ryan_modelling_2023} unless $N>100$ in which case a Gaussian likelihood is used.

Each synthetic VLBI observation produces both a flux and sky coordinates (RA and Dec) with relevant uncertainties.  The flux observation proceeds identically to the radio photometric routine.  The uncertainty $\sigma_c$ in each direction of the centroid is computed as $\sigma_{\rm beam} / \sqrt{\rm SNR}$, where the signal-to-noise ratio SNR is $F_{\rm obs} / \sigma$.  We compute the true RA and Dec from \afterglowpy{} and then add Gaussian noise of width $\sigma_c$ to produce the observed values.  As in the mock photometry, if $F_{\rm obs} < 3 F_{\rm lim}$ the observation is declared a non-detection, a 3$\sigma$ flux upper limit is emitted and the RA and Dec uncertainties are set to large numbers to make the data points unconstraining.  We take the flux uncertainty parameters as above and assign an optimistic $\sigma_{\rm beam} = 1$ mas.  In parameter estimation the flux, RA, and Dec are given a three dimensional multivariate normal likelihood with independent uncertainties (a diagonal covariance matrix).

\begin{figure}
    \centering
    \includegraphics[width=\textwidth]{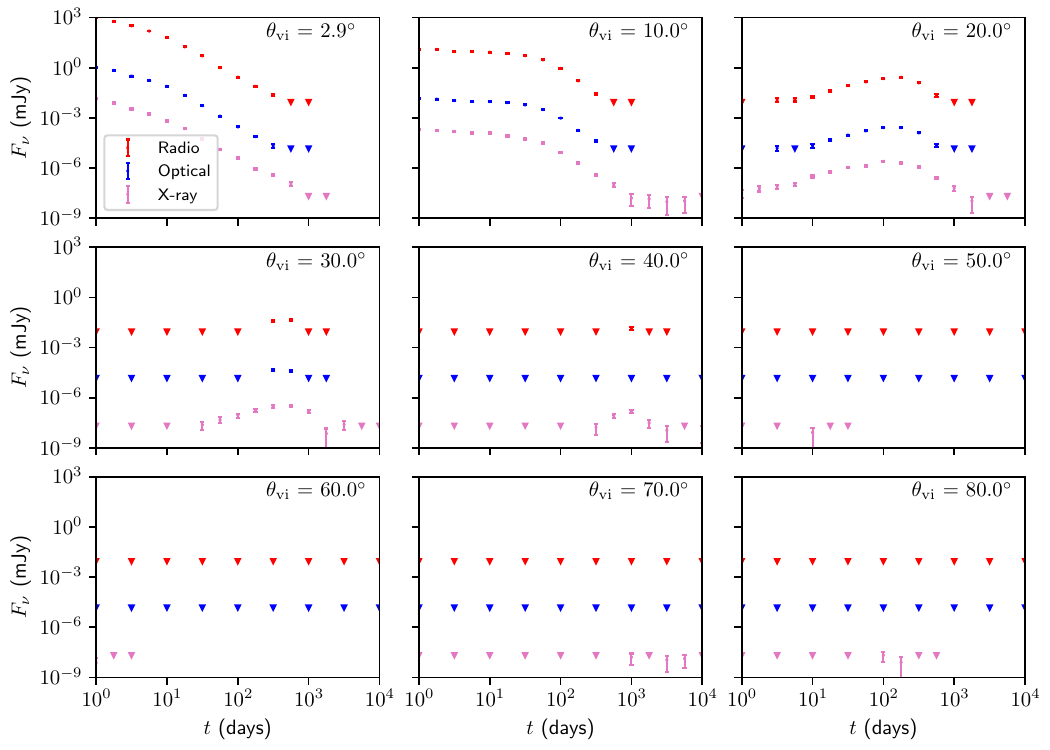}
    \caption{We show the light curve data for the used EM counterparts, generated with a Gaussian jet. The different frequency bands are color coded (red: radio, blue: optical, violet: X-ray), and
    measurements are plotted with an error bar, while upper limits are denoted with an upside-down triangle.}
    \label{fig:mock-lc}
\end{figure}

\gsrfix{For the fitting of the mock data, we employ the following prior ranges in
$\{\log_{10}E_0/{\rm erg}$, $\theta_c$, $\theta_w$, $b$, $\log_{10}n_0/{\rm cm^{-3}}$, $p$, $\log_{10}\epsilon_e$, $\log_{10}\epsilon_B$, $\log_{10}\xi_N$, $\mathrm{RA}_0$, $\mathrm{Dec}_0$, $\mathrm{PA}\}$:
$\mathcal{U}\lp 45,56\rp$, $\mathcal{U}\lp 0.01,\pi/2\rp$ rad,
$\mathcal{U}\lp 0.01,\pi/2\rp$ rad, $\mathcal{U}\lp 0,10\rp$,
$\mathcal{U}\lp -6,10\rp$, $\mathcal{U}\lp 2,5\rp$, 
$\mathcal{U}\lp -6,0\rp$, $\mathcal{U}\lp -6,0\rp$, 
$\mathcal{U}\lp -6,0\rp$, $\mathcal{U}\lp -10,10\rp$ mas, 
$\mathcal{U}\lp -10,10\rp$ mas, $\mathcal{U}\lp 0, 360\rp$ deg. }
Furthermore, the EM inference uses the
same prior for the viewing angle as is used for the inclination in the $\lambda\sgw$-model, i.e. 
$\prio\lb \avi\vert \lambda\sem\rb = \frac{1}{2}\Theta\lb \sin\avi\rb\vert\cos\avi\vert$,
but with range $\avi \in [0,\pi/2]$, and
for the distance prior we use a Gaussian prior in $H_0$ that is centered on a fixed value of
$H_0^{(\text{ref})} = 70$ km s$^{-1}$ Mpc$^{-1}$, i.e. $\prio\lb H_0\vert\lambda\sem\rb = \mathcal{N}\lb \mu = 70, \sigma=10\rb$ 
(in km s$^{-1}$ Mpc$^{-1}$), where $\mathcal{N}\lb \mu,\sigma\rb$ denotes a Gaussian
distribution with mean $\mu$ and variance $\sigma$. This prior
then induces a prior on the luminosity distance, which can be obtained in closed form
if we assume that we are at small redshifts where the luminosity distance is not
sensitive to the specific cosmological model and $H_0 \sim \frac{c z}{\dlum}$, in
which case
$\prio\lb\dlum\vert z,\lambda\sem\rb = \prio\lb H_0\lb\dlum,z\rb\vert\lambda\sem\rb \frac{c z}{\dlum^2}$,
for constant redshift. This is chosen to limit the range in $\dlum$ beyond the $\mathcal{U}\lp 1,100\rp$ Mpc
prior that was specified above, since a large range in $\dlum$ reduces the efficiency of the inference. We
made sure that the precision of the $\dlum$ posteriors from the EM data is always worse than the GW constraints,
such that the joint inference is not affected by this choice.

Fig. \ref{fig:mock-lc} shows the generated mock light curve observations for this work.

\section{Supplemental tables} \label{app:data-supp}

In this section, we summarize results from the different inference campaigns. For every campaign,
we quote the median ($\text{med}\lp H_0\rp$) and the $2.5$th-$97.5$th percentiles of the $H_0$ posteriors
for the different inference scenarios.
Furthermore, we provide statistics that help to estimate the accuracy and precision of the respective
posteriors, we use $\Gamma_{H_0}=\sigma_{H_0}/\text{med}\lp H_0\rp$ to gauge the precision (statistical
error) of the measurement, and since we know the truth in our runs, we define
$\beta_{H_0} = (\text{med}\lp H_0\rp - H_0^{(\text{true})})/ \sigma_{H_0}$ to estimate the accuracy.

To better understand the differences/consistency between different inference models for the EM data, we also define
the quantities 
$\Delta H_0 = (\text{med}\lp H_0\rp_{\text{GJ(c)}} - \text{med}\lp H_0\rp_{\text{PLJ(c)}})/ \text{med}\lp H_0\rp_{\text{GJ(c)}}$, for the relative deviation from the GJ(c) model and
$\delta H_0 = \text{med}\lp H_0\rp_{\text{GJ(c)}} \Delta H_0 / \sigma_{H_0}^{\text{GJ}}$, for 
the deviation from GJ(c) in terms of $\sigma_{H_0}^{\text{GJ}}$, in analogy to our accuracy statistic
$\beta_{H_0}$. We check consistency only among light-curve-only and light curve + centroid motion
inferences respectively and we always use the GJ(c) jet models as reference, since these should also
be consistent with the injection.

We work with unrealistically loud sources, which will suppress the statistical error (estimated in terms
of $\sigma_{H_0}$) to be able to identify systematic errors which might be hidden for sources with higher
statistical errors (there the bias would only show up on the multi-event level, once the statistical error
reaches the bias scale; we choose the sources so close that the statistical error is below the bias scale
that could still be problematic, any bias that we do not resolve at this point would also not become relevant
for the immediate Hubble inference.). This implies that $\Gamma_{H_0}$ and $\Delta H_0$ do not provide
representative properties of the posteriors that would generalize to more realistic sources. Only $\beta_{H_0}$
and $\delta H_0$ which are normalized by the statistical error would provide an idea of how the distributions
might change for larger statistical errors.

In Tab. \ref{tab:no-offset-H0-HPD} and \ref{tab:no-offset-H0-HPD-stats} we show the statistics for the
inference runs that do not use an intrinsic misalignment parameter and where the only inaccuracy can arise
from a biased EM inference.

\begin{table}
    \centering
    \begin{tabular}{c|ccccc}
        \hline
        \hline
        \multirow{2}{*}{$\avi^{(\text{true})}=\avi\lb\tjn^{(\text{true})}\rb$ } 
            & \multicolumn{5}{c}{$H_0$ (median,$2.5$th-$97.5$th percentiles)} \\
            \cline{2-6}
            & GW+z & GW+GJ+z & GW+PLJ+z & GW+GJc+z & GW+PLJc+z \\
        \hline
$3\degree$ & $84.4^{+14.5}_{-13.6}$ & $69.5^{+1.8}_{-2.0}$ & $69.6^{+1.9}_{-2.0}$ & $69.5^{+1.8}_{-2.0}$ & $69.6^{+1.9}_{-1.9}$ \\ 
$10\degree$ & $84.0^{+14.5}_{-14.4}$ & $69.5^{+1.9}_{-1.8}$ & $69.6^{+2.0}_{-1.9}$ & $70.0^{+1.6}_{-1.8}$ & $70.0^{+1.7}_{-1.9}$ \\ 
$20\degree$ & $81.1^{+15.7}_{-15.0}$ & $70.6^{+5.2}_{-4.3}$ & $69.4^{+5.6}_{-3.2}$ & $70.1^{+1.6}_{-1.5}$ & $69.7^{+1.6}_{-1.5}$ \\ 
$30\degree$ & $77.5^{+15.9}_{-16.3}$ & $79.6^{+14.3}_{-12.4}$ & $79.1^{+12.2}_{-11.1}$ & $69.7^{+2.1}_{-2.1}$ & $69.3^{+2.0}_{-2.0}$ \\ 
$40\degree$ & $72.9^{+15.0}_{-17.2}$ & $78.8^{+13.5}_{-15.9}$ & $79.1^{+14.7}_{-14.9}$ & $71.5^{+6.8}_{-5.2}$ & $73.0^{+8.5}_{-6.0}$ \\ 
$50\degree$ & $71.3^{+13.5}_{-19.1}$ & $74.3^{+13.3}_{-16.3}$ & $74.1^{+13.4}_{-16.5}$ & $73.8^{+13.4}_{-16.8}$ & $74.0^{+13.6}_{-17.7}$ \\ 
$60\degree$ & $71.3^{+10.6}_{-14.2}$ & $72.8^{+9.7}_{-12.2}$ & $72.7^{+9.7}_{-11.9}$ & $72.5^{+10.0}_{-12.4}$ & $72.6^{+9.9}_{-12.5}$ \\ 
$70\degree$ & $71.1^{+5.7}_{-7.4}$ & $72.0^{+5.5}_{-6.7}$ & $72.0^{+5.5}_{-6.6}$ & $71.9^{+5.5}_{-6.8}$ & $72.0^{+5.5}_{-7.1}$ \\ 
$80\degree$ & $70.3^{+3.0}_{-3.6}$ & $71.1^{+2.9}_{-3.2}$ & $71.1^{+2.9}_{-3.3}$ & $71.1^{+2.9}_{-3.2}$ & $71.1^{+2.9}_{-3.2}$ \\ 
        \hline
        \hline
    \end{tabular}
    \caption{
        This table lists the posterior results for the Hubble parameter from the $\tmis^{(\text{true})}=0$
        inference campaign for the different inference scenarios
        and all viewing angles, using the median of the posterior distribution and the
        $2.5$th-$97.5$th percentiles.
    }
    \label{tab:no-offset-H0-HPD}
\end{table}

\begin{table}
    \centering
    \begin{tabular}{c|cccccccccc|cccc}
        \hline
        \hline
        \multirow{2}{*}{$\avi^{(\text{true})}$ } 
            & \multicolumn{10}{c}{$H_0$ (median,$2.5$th-$97.5$th percentiles)} & \multicolumn{4}{|c}{posterior consistency} \\
            \cline{2-15}
            & \multicolumn{2}{c}{GW+z} & \multicolumn{2}{c}{GW+GJ+z} & \multicolumn{2}{c}{GW+PLJ+z} & \multicolumn{2}{c}{GW+GJc+z} & \multicolumn{2}{c}{GW+PLJc+z} & \multicolumn{2}{|c}{GJ/PLJ} & \multicolumn{2}{c}{GJc/PLJc} \\
            & $\Gamma_{H_0}$ & $\beta_{H_0}$ & $\Gamma_{H_0}$ & $\beta_{H_0}$ & $\Gamma_{H_0}$ & $\beta_{H_0}$ & $\Gamma_{H_0}$ & $\beta_{H_0}$ & $\Gamma_{H_0}$ & $\beta_{H_0}$ & $\Delta H_0$ & $\delta H_0$ & $\Delta H_0$ & $\delta H_0$ \\
        \hline
   $3\degree$& $0.1$ & $1.8$& $0.01$ & $-0.51$& $0.01$ & $-0.43$& $0.01$ & $-0.53$& $0.01$ & $-0.4$& $-0.0$ & $-0.07$& $-0.0$ & $-0.12$ \\ 
$10\degree$& $0.1$ & $1.69$& $0.01$ & $-0.53$& $0.02$ & $-0.36$& $0.01$ & $0.03$& $0.01$ & $0.03$& $-0.0$ & $-0.07$& $-0.0$ & $-0.01$ \\ 
$20\degree$& $0.11$ & $1.29$& $0.04$ & $0.24$& $0.03$ & $-0.26$& $0.01$ & $0.13$& $0.01$ & $-0.33$& $0.02$ & $0.49$& $0.01$ & $0.46$ \\ 
$30\degree$& $0.12$ & $0.84$& $0.09$ & $1.39$& $0.08$ & $1.45$& $0.02$ & $-0.32$& $0.01$ & $-0.65$& $0.01$ & $0.08$& $0.0$ & $0.3$ \\ 
$40\degree$& $0.12$ & $0.33$& $0.1$ & $1.17$& $0.1$ & $1.21$& $0.04$ & $0.48$& $0.05$ & $0.77$& $-0.0$ & $-0.03$& $-0.02$ & $-0.45$ \\ 
$50\degree$& $0.12$ & $0.15$& $0.1$ & $0.57$& $0.1$ & $0.55$& $0.1$ & $0.5$& $0.11$ & $0.51$& $0.0$ & $0.02$& $-0.0$ & $-0.03$ \\ 
$60\degree$& $0.09$ & $0.2$& $0.08$ & $0.49$& $0.08$ & $0.49$& $0.08$ & $0.43$& $0.08$ & $0.45$& $0.0$ & $0.01$& $-0.0$ & $-0.03$ \\ 
$70\degree$& $0.05$ & $0.33$& $0.04$ & $0.62$& $0.04$ & $0.63$& $0.04$ & $0.6$& $0.05$ & $0.61$& $-0.0$ & $-0.01$& $-0.0$ & $-0.02$ \\ 
$80\degree$& $0.02$ & $0.18$& $0.02$ & $0.7$& $0.02$ & $0.72$& $0.02$ & $0.72$& $0.02$ & $0.74$& $-0.0$ & $-0.03$& $-0.0$ & $-0.02$ \\ 
        \hline
        \hline
    \end{tabular}
    \caption{
        This table lists statistics for the Hubble posterior from the $\tmis^{(\text{true})}=0$
        inference campaign for the different inference scenarios
        and all viewing angles, with two decimal places($0.0$ means a number $<0.01$). We
        quote $\Gamma_{H_0}=\sigma_{H_0}/\text{med}\lp H_0\rp$ (for precision), $\beta_{H_0} = (\text{med}\lp H_0\rp - H_0^{(\text{true})})/ \sigma_{H_0}$ (for accuracy), $\Delta H_0 = (\text{med}\lp H_0\rp_{\text{GJ(c)}} - \text{med}\lp H_0\rp_{\text{PLJ(c)}})/ \text{med}\lp H_0\rp_{\text{GJ(c)}}$ (for relative deviation from GJ(c) model) and
        $\delta H_0 = \text{med}\lp H_0\rp_{\text{GJ(c)}} \Delta H_0 / \sigma_{H_0}^{\text{GJ}}$ (for deviation from GJ(c) in terms of $\sigma_{H_0}^{\text{GJ}}$).
    }
    \label{tab:no-offset-H0-HPD-stats}
\end{table}

In Tab. \ref{tab:offsetGrid} and \ref{tab:offsetGrid-stats} we show the statistics for the
$H_0$-posteriors in Fig. \ref{fig:offsetGrid}, which varies the intrinsic misalignment
systematically around the aligned case for low to intermediate $\avi$.

\begin{table}
    \centering
    \begin{tabular}{cc|cccc}
        \hline
        \hline
        \multirow{2}{*}{$\avi^{(\text{true})}$ } & \multirow{2}{*}{$\tmis^{(\text{true})}=\avi\lb\tjn^{(\text{true})}\rb - \avi^{(\text{true})}$ } & \multicolumn{4}{c}{$H_0$ (median,$2.5$th-$97.5$th percentiles)} \\
            \cline{3-6}
            & & GW+GJ+z & GW+PLJ+z & GW+GJc+z & GW+PLJc+z \\
        \hline
        \multirow{7}{*}{$20\degree$ }
& $-6\degree$ & $72.7^{+5.3}_{-4.5}$ & $71.4^{+5.5}_{-3.2}$ & $72.5^{+1.6}_{-1.6}$ & $72.1^{+1.6}_{-1.6}$ \\ 
& $-3\degree$ & $71.7^{+5.3}_{-4.4}$ & $70.5^{+5.6}_{-3.3}$ & $71.4^{+1.6}_{-1.5}$ & $71.0^{+1.6}_{-1.5}$ \\ 
& $0\degree$ & $70.6^{+5.2}_{-4.3}$ & $69.4^{+5.6}_{-3.2}$ & $70.1^{+1.6}_{-1.5}$ & $69.7^{+1.6}_{-1.5}$ \\ 
& $3\degree$ & $68.9^{+5.1}_{-4.2}$ & $67.8^{+5.6}_{-3.1}$ & $68.4^{+1.5}_{-1.5}$ & $68.0^{+1.5}_{-1.5}$ \\ 
& $6\degree$ & $67.7^{+5.0}_{-4.0}$ & $66.7^{+5.7}_{-3.1}$ & $67.0^{+1.5}_{-1.4}$ & $66.7^{+1.5}_{-1.5}$ \\ 
& $10\degree$ & $65.4^{+4.8}_{-3.9}$ & $64.6^{+5.7}_{-3.1}$ & $64.5^{+1.6}_{-1.5}$ & $64.3^{+1.6}_{-1.5}$ \\ 
& $14\degree$ & $62.9^{+4.5}_{-3.9}$ & $62.0^{+5.6}_{-3.0}$ & $61.7^{+1.4}_{-1.5}$ & $61.5^{+1.4}_{-1.5}$ \\ 
        \hline
        \multirow{10}{*}{$30\degree$ }
& $-16\degree$ & $86.8^{+13.2}_{-11.0}$ & $86.1^{+11.9}_{-10.4}$ & $80.0^{+3.1}_{-2.7}$ & $79.4^{+2.7}_{-2.5}$ \\ 
& $-13\degree$ & $85.4^{+14.0}_{-11.1}$ & $84.7^{+12.5}_{-10.2}$ & $78.5^{+2.9}_{-2.6}$ & $77.9^{+2.6}_{-2.4}$ \\ 
& $-10\degree$ & $84.3^{+14.3}_{-11.3}$ & $83.5^{+12.7}_{-10.2}$ & $76.8^{+2.8}_{-2.5}$ & $76.3^{+2.5}_{-2.3}$ \\ 
& $-7\degree$ & $83.0^{+14.3}_{-11.8}$ & $82.2^{+12.6}_{-10.6}$ & $74.5^{+2.5}_{-2.3}$ & $74.1^{+2.3}_{-2.2}$ \\ 
& $-4\degree$ & $81.6^{+14.8}_{-11.7}$ & $80.9^{+13.0}_{-10.5}$ & $72.9^{+2.4}_{-2.3}$ & $72.5^{+2.3}_{-2.2}$ \\ 
& $0\degree$ & $79.6^{+14.3}_{-12.4}$ & $79.1^{+12.2}_{-11.1}$ & $69.7^{+2.1}_{-2.1}$ & $69.3^{+2.0}_{-2.0}$ \\ 
& $4\degree$ & $76.6^{+13.8}_{-12.0}$ & $75.9^{+11.9}_{-11.0}$ & $66.1^{+1.9}_{-1.8}$ & $65.8^{+1.9}_{-1.8}$ \\ 
& $7\degree$ & $75.8^{+13.9}_{-12.2}$ & $74.8^{+12.6}_{-11.4}$ & $64.1^{+1.9}_{-1.8}$ & $63.9^{+1.9}_{-1.8}$ \\ 
& $10\degree$ & $73.5^{+13.9}_{-11.9}$ & $72.7^{+12.9}_{-10.9}$ & $61.5^{+1.8}_{-1.8}$ & $61.3^{+1.7}_{-1.7}$ \\ 
& $14\degree$ & $71.4^{+14.5}_{-12.3}$ & $70.5^{+13.8}_{-11.2}$ & $58.0^{+1.6}_{-1.6}$ & $58.0^{+1.6}_{-1.6}$ \\ 
        \hline
        \multirow{11}{*}{$40\degree$ }
& $-17\degree$ & $86.9^{+13.1}_{-13.9}$ & $87.5^{+12.5}_{-14.6}$ & $92.0^{+7.5}_{-6.8}$ & $93.0^{+7.0}_{-6.2}$ \\ 
& $-14\degree$ & $86.0^{+14.0}_{-14.4}$ & $86.6^{+13.4}_{-15.9}$ & $89.9^{+7.9}_{-7.5}$ & $91.1^{+7.7}_{-7.2}$ \\ 
& $-10\degree$ & $84.3^{+14.3}_{-15.9}$ & $84.5^{+14.2}_{-16.0}$ & $84.6^{+8.2}_{-7.1}$ & $86.1^{+8.7}_{-7.4}$ \\ 
& $-6\degree$ & $81.2^{+13.6}_{-21.8}$ & $81.2^{+14.7}_{-16.0}$ & $78.9^{+7.8}_{-6.4}$ & $80.5^{+8.6}_{-7.1}$ \\ 
& $-3\degree$ & $80.8^{+13.5}_{-17.6}$ & $80.9^{+14.8}_{-15.8}$ & $75.9^{+7.4}_{-5.9}$ & $77.5^{+8.6}_{-6.9}$ \\ 
& $0\degree$ & $78.8^{+13.5}_{-15.9}$ & $79.1^{+14.7}_{-14.9}$ & $71.5^{+6.8}_{-5.2}$ & $73.0^{+8.5}_{-6.0}$ \\ 
& $4\degree$ & $77.6^{+13.3}_{-18.6}$ & $77.5^{+13.9}_{-15.1}$ & $66.3^{+5.6}_{-4.4}$ & $67.8^{+9.9}_{-5.5}$ \\ 
& $7\degree$ & $76.4^{+12.9}_{-17.9}$ & $76.5^{+13.4}_{-15.2}$ & $62.9^{+4.8}_{-3.9}$ & $64.1^{+13.7}_{-5.0}$ \\ 
& $10\degree$ & $75.1^{+12.6}_{-15.6}$ & $75.3^{+13.0}_{-14.7}$ & $59.5^{+4.2}_{-3.6}$ & $60.9^{+16.0}_{-4.7}$ \\ 
& $13\degree$ & $74.4^{+11.4}_{-14.7}$ & $74.9^{+11.4}_{-14.4}$ & $56.9^{+3.6}_{-3.3}$ & $58.4^{+16.4}_{-4.6}$ \\ 
& $15\degree$ & $73.6^{+11.2}_{-13.9}$ & $74.3^{+10.8}_{-13.7}$ & $54.4^{+3.7}_{-3.0}$ & $56.7^{+22.6}_{-4.8}$ \\ 
        \hline
        \hline
    \end{tabular}
    \caption{
        This table lists the median and the $2.5$th-$97.5$th percentiles of the posterior
        results for the Hubble parameter from the $\tmis^{(\text{true})}\neq0$ datasets at varying
        $\avi^{(\text{true})}$, shown in Fig. \ref{fig:offsetGrid}.
    }
    \label{tab:offsetGrid}
\end{table}

\begin{table}
    \centering
    \begin{tabular}{cc|cccccccc|cccc}
        \hline
        \hline
        \multirow{3}{*}{$\avi^{(\text{true})}$ } & \multirow{3}{*}{$\tmis^{(\text{true})}$ } 
            & \multicolumn{8}{c}{$H_0$ posterior statistics} & \multicolumn{4}{|c}{posterior consistency} \\
            \cline{3-14}
            & & \multicolumn{2}{c}{GW+GJ+z} & \multicolumn{2}{c}{GW+PLJ+z} & \multicolumn{2}{c}{GW+GJc+z} & \multicolumn{2}{c}{GW+PLJc+z} & \multicolumn{2}{|c}{GJ/PLJ} & \multicolumn{2}{c}{GJc/PLJc} \\
            & & $\Gamma_{H_0}$ & $\beta_{H_0}$ & $\Gamma_{H_0}$ & $\beta_{H_0}$ & $\Gamma_{H_0}$ & $\beta_{H_0}$ & $\Gamma_{H_0}$ & $\beta_{H_0}$ & $\Delta H_0$ & $\delta H_0$ & $\Delta H_0$ & $\delta H_0$ \\
            \hline
            \multirow{7}{*}{$20\degree$ }
& $-6\degree$& $0.04$ & $1.05$& $0.03$ & $0.62$& $0.01$ & $3.02$& $0.01$ & $2.52$& $0.02$ & $0.49$& $0.01$ & $0.49$ \\ 
& $-3\degree$& $0.04$ & $0.69$& $0.03$ & $0.22$& $0.01$ & $1.73$& $0.01$ & $1.25$& $0.02$ & $0.49$& $0.01$ & $0.47$ \\ 
& $0\degree$& $0.04$ & $0.24$& $0.03$ & $-0.26$& $0.01$ & $0.13$& $0.01$ & $-0.33$& $0.02$ & $0.49$& $0.01$ & $0.46$ \\ 
& $3\degree$& $0.04$ & $-0.43$& $0.03$ & $-0.94$& $0.01$ & $-2.11$& $0.01$ & $-2.53$& $0.02$ & $0.48$& $0.0$ & $0.43$ \\ 
& $6\degree$& $0.03$ & $-0.96$& $0.04$ & $-1.39$& $0.01$ & $-3.93$& $0.01$ & $-4.33$& $0.01$ & $0.42$& $0.0$ & $0.42$ \\ 
& $10\degree$& $0.03$ & $-2.03$& $0.04$ & $-2.26$& $0.01$ & $-6.8$& $0.01$ & $-7.16$& $0.01$ & $0.37$& $0.0$ & $0.33$ \\ 
& $14\degree$& $0.03$ & $-3.29$& $0.04$ & $-3.4$& $0.01$ & $-11.18$& $0.01$ & $-11.53$& $0.01$ & $0.38$& $0.0$ & $0.34$ \\ 
        \hline
        \multirow{10}{*}{$30\degree$ }
 & $-16\degree$& $0.08$ & $2.53$& $0.07$ & $2.68$& $0.02$ & $6.67$& $0.02$ & $6.97$& $0.01$ & $0.11$& $0.01$ & $0.41$ \\ 
& $-13\degree$& $0.08$ & $2.28$& $0.07$ & $2.41$& $0.02$ & $5.95$& $0.02$ & $6.13$& $0.01$ & $0.11$& $0.01$ & $0.4$ \\ 
& $-10\degree$& $0.08$ & $2.1$& $0.07$ & $2.21$& $0.02$ & $5.06$& $0.02$ & $5.1$& $0.01$ & $0.12$& $0.01$ & $0.39$ \\ 
& $-7\degree$& $0.08$ & $1.89$& $0.07$ & $2.0$& $0.02$ & $3.64$& $0.02$ & $3.51$& $0.01$ & $0.11$& $0.01$ & $0.36$ \\ 
& $-4\degree$& $0.08$ & $1.68$& $0.08$ & $1.77$& $0.02$ & $2.4$& $0.02$ & $2.17$& $0.01$ & $0.1$& $0.01$ & $0.35$ \\ 
& $0\degree$& $0.09$ & $1.39$& $0.08$ & $1.45$& $0.02$ & $-0.32$& $0.01$ & $-0.65$& $0.01$ & $0.08$& $0.0$ & $0.3$ \\ 
& $4\degree$& $0.09$ & $0.97$& $0.08$ & $0.95$& $0.01$ & $-4.08$& $0.01$ & $-4.47$& $0.01$ & $0.1$& $0.0$ & $0.29$ \\ 
& $7\degree$& $0.09$ & $0.85$& $0.09$ & $0.75$& $0.01$ & $-6.18$& $0.01$ & $-6.53$& $0.01$ & $0.14$& $0.0$ & $0.22$ \\ 
& $10\degree$& $0.09$ & $0.52$& $0.09$ & $0.42$& $0.01$ & $-9.31$& $0.01$ & $-9.7$& $0.01$ & $0.13$& $0.0$ & $0.16$ \\ 
& $14\degree$& $0.1$ & $0.2$& $0.09$ & $0.08$& $0.01$ & $-14.51$& $0.01$ & $-14.62$& $0.01$ & $0.12$& $0.0$ & $0.01$ \\ 
        \hline
        \multirow{11}{*}{$40\degree$ }
& $-17\degree$& $0.09$ & $2.23$& $0.09$ & $2.33$& $0.04$ & $5.67$& $0.04$ & $6.26$& $-0.01$ & $-0.08$& $-0.01$ & $-0.26$ \\ 
& $-14\degree$& $0.09$ & $2.05$& $0.09$ & $2.16$& $0.05$ & $4.82$& $0.04$ & $5.4$& $-0.01$ & $-0.08$& $-0.01$ & $-0.3$ \\ 
& $-10\degree$& $0.09$ & $1.82$& $0.09$ & $1.84$& $0.05$ & $3.62$& $0.05$ & $3.92$& $-0.0$ & $-0.02$& $-0.02$ & $-0.35$ \\ 
& $-6\degree$& $0.1$ & $1.38$& $0.1$ & $1.45$& $0.05$ & $2.36$& $0.05$ & $2.58$& $-0.0$ & $-0.0$& $-0.02$ & $-0.42$ \\ 
& $-3\degree$& $0.1$ & $1.36$& $0.1$ & $1.41$& $0.05$ & $1.67$& $0.05$ & $1.86$& $-0.0$ & $-0.01$& $-0.02$ & $-0.45$ \\ 
& $0\degree$& $0.1$ & $1.17$& $0.1$ & $1.21$& $0.04$ & $0.48$& $0.05$ & $0.77$& $-0.0$ & $-0.03$& $-0.02$ & $-0.45$ \\ 
& $4\degree$& $0.1$ & $1.01$& $0.1$ & $1.01$& $0.04$ & $-1.36$& $0.06$ & $-0.57$& $0.0$ & $0.01$& $-0.02$ & $-0.53$ \\ 
& $7\degree$& $0.1$ & $0.86$& $0.1$ & $0.87$& $0.04$ & $-2.95$& $0.07$ & $-1.39$& $-0.0$ & $-0.02$& $-0.02$ & $-0.52$ \\ 
& $10\degree$& $0.09$ & $0.73$& $0.09$ & $0.75$& $0.04$ & $-4.09$& $0.09$ & $-1.6$& $-0.0$ & $-0.03$& $-0.02$ & $-0.54$ \\ 
& $13\degree$& $0.09$ & $0.67$& $0.09$ & $0.74$& $0.05$ & $-4.88$& $0.11$ & $-1.89$& $-0.01$ & $-0.07$& $-0.03$ & $-0.56$ \\ 
& $15\degree$& $0.09$ & $0.57$& $0.08$ & $0.7$& $0.06$ & $-4.59$& $0.12$ & $-1.96$& $-0.01$ & $-0.11$& $-0.04$ & $-0.69$ \\ 
        \hline
        \hline
    \end{tabular}
    \caption{
        This table lists statistics for the Hubble posterior from the
        $\tmis^{(\text{true})}\neq0$ datasets at varying $\avi^{(\text{true})}$, shown
        in Fig. \ref{fig:offsetGrid},
        for the different inference campaigns
        and all viewing angles, with two decimal places($0.0$ means a number $<0.01$). We
        quote $\Gamma_{H_0}=\sigma_{H_0}/\text{med}\lp H_0\rp$ (for precision), $\beta_{H_0} = (\text{med}\lp H_0\rp - H_0^{(\text{true})})/ \sigma_{H_0}$ (for accuracy), $\Delta H_0 = (\text{med}\lp H_0\rp_{\text{GJ(c)}} - \text{med}\lp H_0\rp_{\text{PLJ(c)}})/ \text{med}\lp H_0\rp_{\text{GJ(c)}}$ (for relative deviation from GJ(c) model) and
        $\delta H_0 = \text{med}\lp H_0\rp_{\text{GJ(c)}} \Delta H_0 / \sigma_{H_0}^{\text{GJ}}$ (for deviation from GJ(c) in terms of $\sigma_{H_0}^{\text{GJ}}$).
    }
    \label{tab:offsetGrid-stats}
\end{table}

\begin{figure*}[ht]
    \begin{center}
        \includegraphics[trim=1cm 0.72cm 1cm 0,clip,scale=0.58]{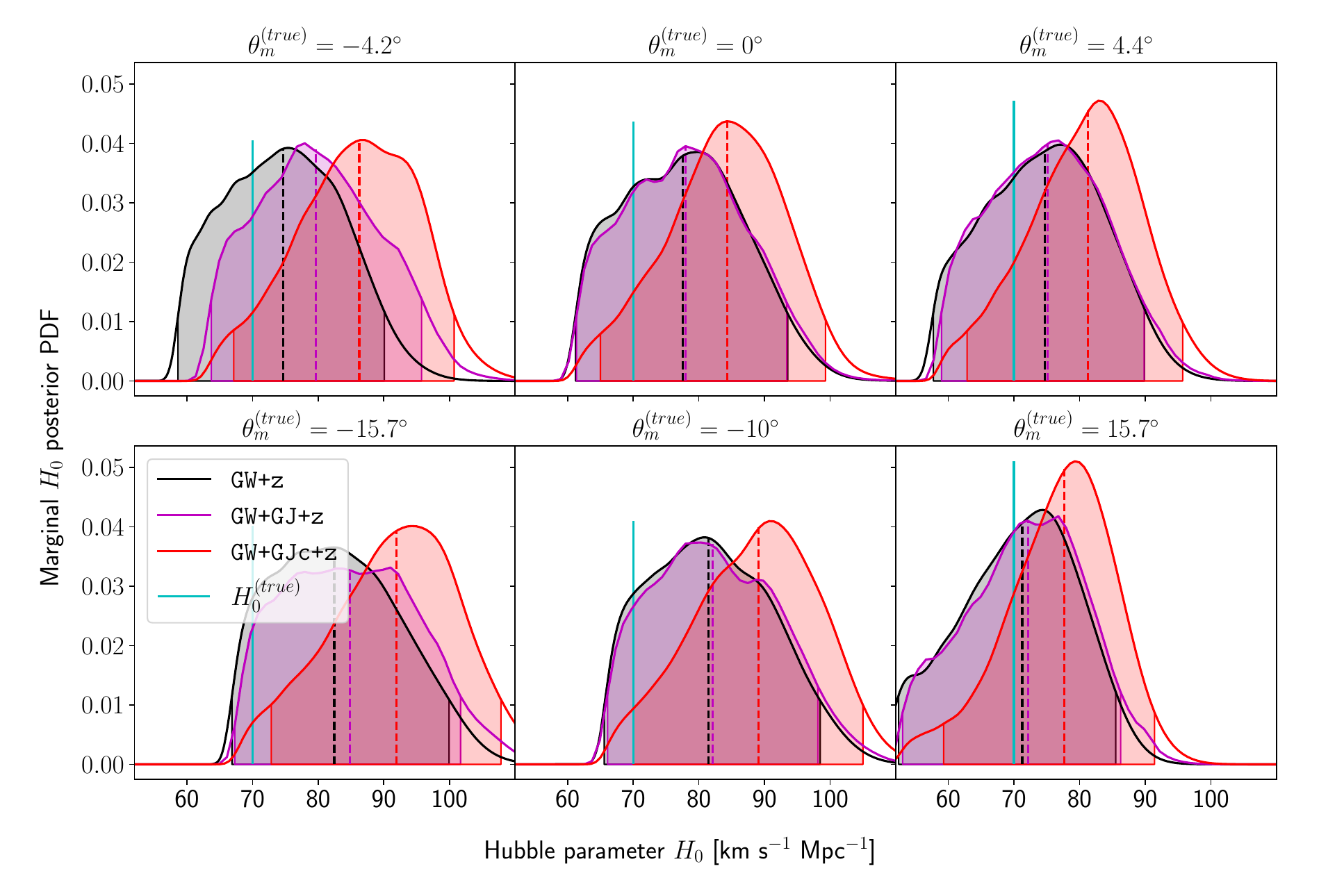}
    \end{center}
    \caption{
    \mmfix{
        This figure shows the marginal posterior PDFs for $H_0$
        from the $\tmis^{(\text{true})}\neq0$ datasets at 
        $\avi^{(\text{true})}=30\degree$, in the case of the co-inference of $(\tmis,H_0)$, as discussed in Sec. \ref{sec:mitigation}. We present the same angles as in Fig. 
        \ref{fig:mitigationGrid} and        
        the median and the $2.5$th-$97.5$th percentiles of the marginal posterior
        are shown with a dashed line and colored regions (using the same color as for the PDF)
        respectively. 
    }
    }
    \label{fig:marginal-mitigationGrid}
\end{figure*}
\begin{figure*}[ht]
    \begin{center}
        \includegraphics[trim=1cm 0.4cm 1cm 0,clip,scale=0.58]{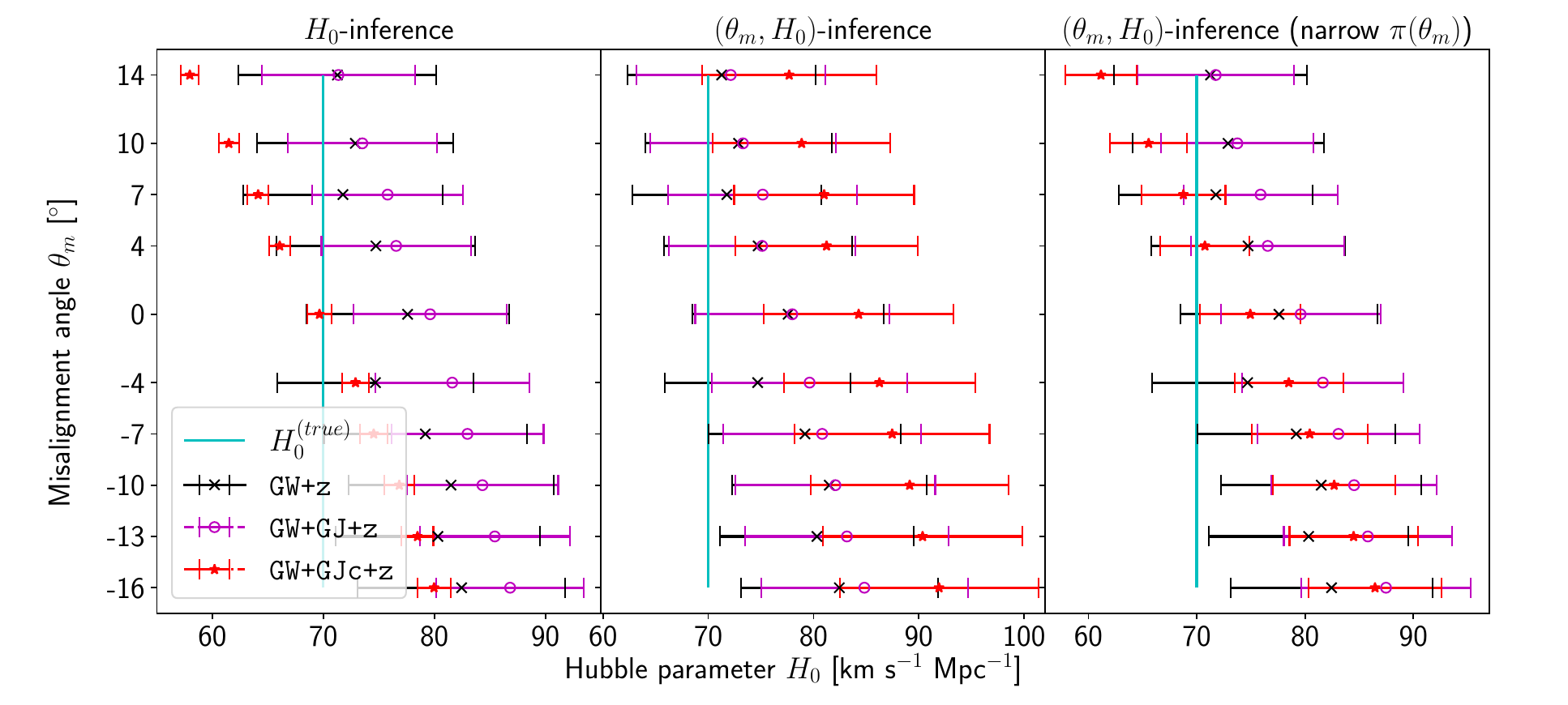}
    \end{center}
    \caption{
    \mmfix{
        This figure shows the median and standard deviation (with error bars) for the
        (marginal) $H_0$ posteriors, considering all available 
        $\tmis^{(\text{true})}\neq0$ datasets at $\avi^{(\text{true})}=30\degree$. The
        left most set shows the behavior for standard inference with $H_0$, and assuming
        $\tmis^{(\text{true})}=0$, and complements Fig. \ref{fig:offset-G4EM3-EM9} and
        Fig. \ref{fig:offsetGrid}. The set in the middle refers to 
        the co-inference of $(\tmis,H_0)$, as discussed in Sec. \ref{sec:mitigation},
        with a $\mathcal{U}\lb -60\degree,60\degree\rb$ prior for $\tmis$ (complementing
        Fig. \ref{fig:mitigationGrid} and \ref{fig:marginal-mitigationGrid}). And the set
        on the right also considers the co-inference of $(\tmis,H_0)$, but with a narrower
        prior of $\mathcal{U}\lb -10\degree,10\degree\rb$ prior for $\tmis$, which complements
        Fig. \ref{fig:marginal-mitigationGrid-narrow-tmis}.
        } 
    }
    \label{fig:comparison-stats-inference-modes}
\end{figure*}
\begin{figure*}[ht]
    \begin{center}
        \includegraphics[trim=1cm 0.72cm 1cm 0,clip,scale=0.58]{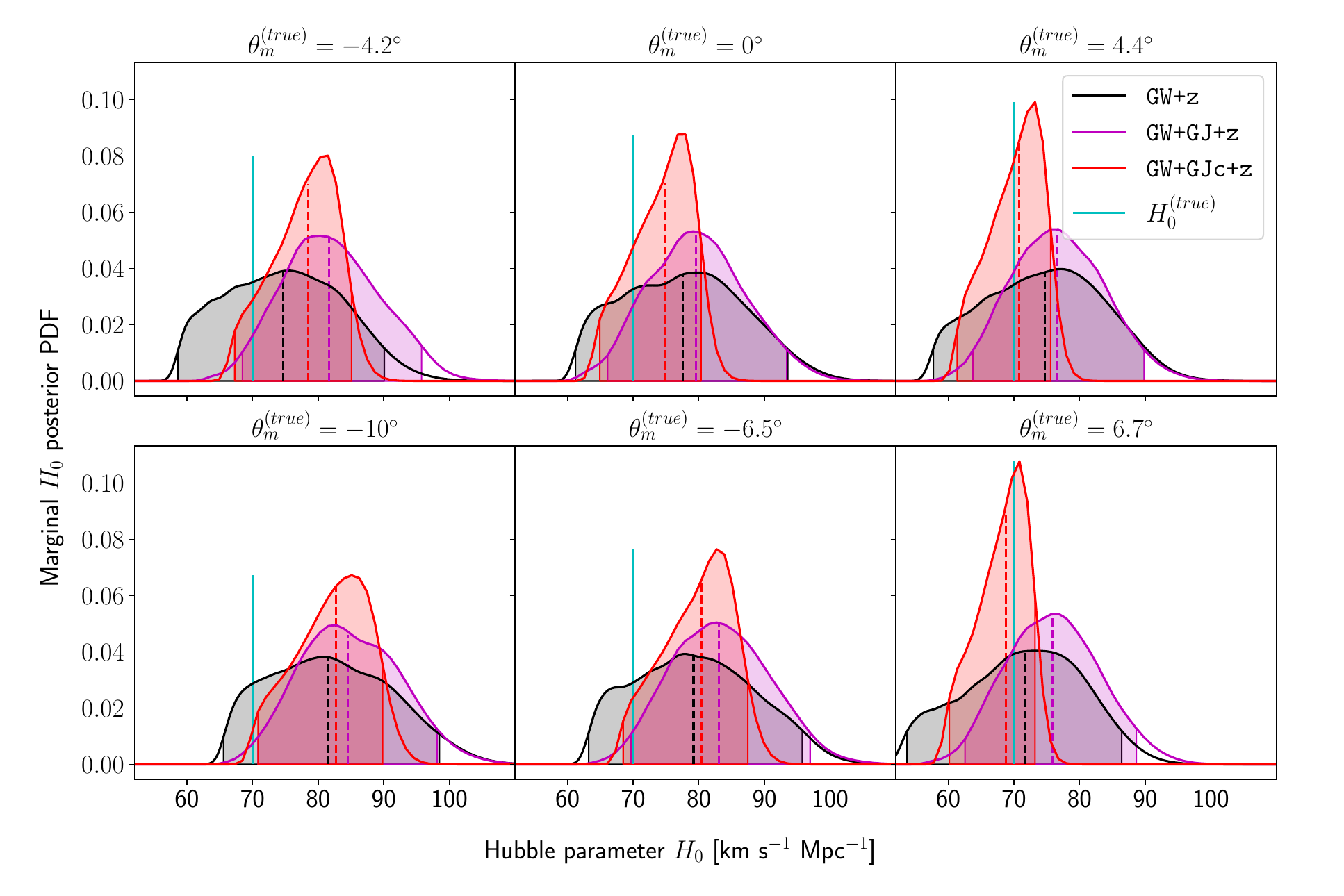}
    \end{center}
    \caption{
    \mmfix{
        This figure shows the marginal posterior PDFs for $H_0$
        from the $\tmis^{(\text{true})}\neq0$ datasets at 
        $\avi^{(\text{true})}=30\degree$, in the case of the co-inference of $(\tmis,H_0)$,
        using the narrower prior $\mathcal{U}\lb-10\degree,10\degree\rb$ for $\tmis$, as discussed in Sec. \ref{sec:mitigation}. 
        We present the data for $\tmis^{(\text{true})} \in\lc -10\degree,-6.5\degree,-4.2\degree,0\degree,4.2\degree,6.7\degree\rc$ and        
        the median and the $2.5$th-$97.5$th percentiles of the marginal posterior
        are shown with a dashed line and colored regions (using the same color as for the PDF)
        respectively. 
    }
    }
    \label{fig:marginal-mitigationGrid-narrow-tmis}
\end{figure*}

\mmfix{Fig. \ref{fig:marginal-mitigationGrid} and column 2 in Fig. \ref{fig:comparison-stats-inference-modes}} 
show the statistics for the marginal $H_0$-posteriors 
\mmfix{from the bias mitigating inference of $H_0$ and $\tmis$ with a
subsequent marginalization over $\tmis$ (we shall hereafter refer to this as $\lb\tmis,H_0\rb$-inference). In particular, we consider scenarios where we combine
$\avi^{(\text{true})}=30\degree$ with varying values of $\tmis^{(\text{true})}\neq0$, which
are also shown in Fig. \ref{fig:mitigationGrid}). This complements the previous inferences
with $\tmis^{(\text{true})}\neq0$, where we assumed $\tmis^{(\text{true})}=0$, and obtained
the biased posteriors in Fig. \ref{fig:offset-G4EM3-EM9} and \ref{fig:offsetGrid}.
}
\mmfix{For comparison, in Fig. \ref{fig:marginal-mitigationGrid-narrow-tmis} and column 3 in Fig. \ref{fig:comparison-stats-inference-modes} 
we present
statistics for the same analysis, when we employ a narrower prior on the value of $\tmis$,
in particular, we use $\prio\lb\tmis\vert \lamH\rb = \mathcal{U}\lb-10\degree,10\degree\rb$.}

\begin{figure*}[ht]
    \begin{center}
        \includegraphics[trim=1cm 0.4cm 1cm 0,clip,scale=0.58]{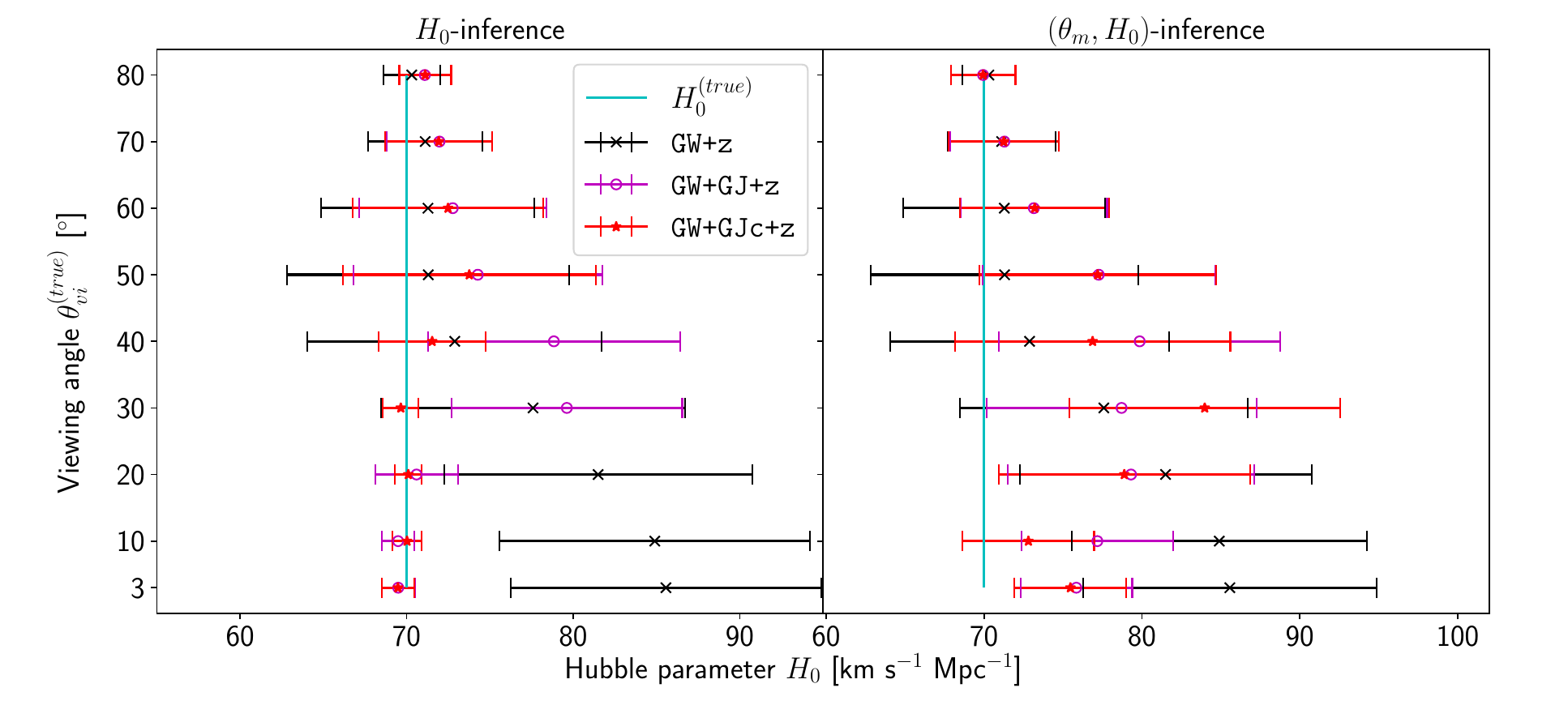}
    \end{center}
    \caption{
    \mmfix{
        This figure shows the median and standard deviation (with error bars) for the
        (marginal) $H_0$ posteriors, considering all available 
        $\tmis^{(\text{true})}=$ datasets. The
        left column shows the behavior for standard inference with $H_0$, which assumes
        $\tmis^{(\text{true})}=0$, The right column refers to 
        the co-inference of $(\tmis,H_0)$, as discussed in Sec. \ref{sec:mitigation},
        with a $\mathcal{U}\lb -60\degree,60\degree\rb$ prior for $\tmis$.
        } 
    }
    \label{fig:tmisZero-comparison-stats-inference-modes}
\end{figure*}

Tab. \ref{tab:tmisZeroH0} and \mmfix{Fig. \ref{fig:tmisZero-comparison-stats-inference-modes}}
show the statistics corresponding
to the marginalized $H_0$ posteriors from the bias-mitigating
inference of all aligned ($\tmis^{(\text{true})}=0$) datasets. Based on these one can judge
if this method can recover zero misalignment angle for the aligned GW and EM datasets.

\mmfix{The presentations in Fig. \ref{fig:comparison-stats-inference-modes} and \ref{fig:tmisZero-comparison-stats-inference-modes} should
allow for an easy extraction of $\text{med}\lp H_0\rp$ and $\sigma_{H_0}$ and therefore
$\beta_{H_0}$ and $\Gamma_{H_0}$.}

\begin{table}
    \centering
    \begin{tabular}{c|cccc}
        \hline
        \hline
        \multirow{2}{*}{$\avi^{(\text{true})}=\avi\lb\tjn^{(\text{true})}\rb$ } 
            & \multicolumn{4}{c}{$H_0$ (median,$2.5$th-$97.5$th percentiles)} \\
            \cline{2-5}
            & GW+GJ+z & GW+PLJ+z & GW+GJc+z & GW+PLJc+z \\
        \hline
 $3\degree$ & $75.8^{+6.2}_{-6.9}$ & $76.1^{+6.3}_{-7.2}$ & $75.5^{+6.3}_{-6.8}$ & $75.3^{+6.8}_{-6.9}$ \\ 
 $10\degree$ & $77.2^{+8.1}_{-8.9}$ & $77.8^{+8.4}_{-10.1}$ & $72.8^{+9.3}_{-5.8}$ & $78.5^{+14.4}_{-11.4}$ \\ 
 $20\degree$ & $79.3^{+14.2}_{-14.1}$ & $77.7^{+12.4}_{-12.7}$ & $78.9^{+15.0}_{-13.9}$ & $73.0^{+17.1}_{-9.2}$ \\ 
 $30\degree$ & $78.7^{+15.9}_{-15.9}$ & $79.1^{+15.3}_{-16.4}$ & $84.0^{+15.4}_{-16.9}$ & $81.1^{+15.1}_{-17.7}$ \\ 
 $40\degree$ & $79.9^{+15.6}_{-17.6}$ & $80.5^{+15.2}_{-16.4}$ & $76.9^{+15.4}_{-18.3}$ & $76.9^{+15.2}_{-18.1}$ \\ 
 $50\degree$ & $77.3^{+11.3}_{-16.7}$ & $77.4^{+11.4}_{-17.1}$ & $77.2^{+11.6}_{-17.0}$ & $76.9^{+11.8}_{-17.2}$ \\ 
 $60\degree$ & $73.2^{+9.4}_{-9.8}$ & $73.2^{+9.4}_{-9.7}$ & $73.2^{+9.3}_{-10.2}$ & $73.2^{+9.4}_{-10.3}$ \\ 
 $70\degree$ & $71.3^{+5.7}_{-7.6}$ & $71.3^{+5.6}_{-7.6}$ & $71.3^{+5.8}_{-7.7}$ & $71.3^{+5.4}_{-7.6}$ \\ 
 $80\degree$ & $69.9^{+4.0}_{-4.6}$ & $69.9^{+4.1}_{-4.7}$ & $69.9^{+4.1}_{-4.6}$ & $69.9^{+4.1}_{-4.6}$ \\ 
        \hline
        \hline
    \end{tabular}
    \caption{
        This table lists the median and the $2.5$th-$97.5$th percentiles of the marginal posterior
        results for the Hubble parameter from the $\tmis^{(\text{true})}=0$ datasets, in the case of
        the co-inference of $(\tmis,H_0)$, as discussed in Sec. \ref{sec:mitigation}.
    }
    \label{tab:tmisZeroH0}
\end{table}

\bibliography{bibliography}{}
\bibliographystyle{aasjournal}
\end{document}